\documentclass[a4paper,fleqn,usenatbib,useAMS]{mnras}

\usepackage{graphicx}	
\usepackage{amsmath}	
\usepackage{amssymb}	
\usepackage{multicol}        
\usepackage{bm}		
\usepackage{pdflscape}	
\usepackage{multirow}
\usepackage{xcolor}
\usepackage{hyperref}
\usepackage{booktabs}
\usepackage{subfigure}
\usepackage{makecell}
\usepackage{tikz}
\usepackage[normalem]{ulem}

\newcommand{\kms}{\,km\,s$^{-1}$} 
\def\specialname[#1]{\textbf{\textsc{#1}}}

\definecolor{lime}{HTML}{A6CE39}
\DeclareRobustCommand{\orcidicon}{%
	\begin{tikzpicture}
	\draw[lime, fill=lime] (0,0)
	circle [radius=0.16]
	node[white] {{\fontfamily{qag}\selectfont \tiny ID}};
	\draw[white, fill=white] (-0.0625,0.095)
	circle [radius=0.007];
	\end{tikzpicture}
	\hspace{-2mm}
}
\foreach \x in {A, ..., Z}{%
	\expandafter\xdef\csname orcid\x\endcsname{\noexpand\href{https://orcid.org/\csname orcidauthor\x\endcsname}{\noexpand\orcidicon}}
}

\usepackage[T1]{fontenc}
\usepackage{ae,aecompl}

\usepackage{newtxtext,newtxmath}

\title[Characterizing the assembly of dark matter halos with protohalo size histories]{
Characterizing the assembly of dark matter halos with protohalo size
    histories: I. Redshift evolution, relation to descendant halos, and halo
assembly bias}
\author[Wang et al.]{Kai Wang\orcidK{},$^{1}$\thanks{Contact e-mail:
    wkcosmology@gmail.com}
	H.J. Mo,$^{2}$
    Yangyao Chen\orcidC{},$^{3, 4}$
    Huiyuan Wang\orcidW{},$^{3, 4}$
	Xiaohu Yang,$^{5, 6, 7, 8}$
    \newauthor
	Jiaqi Wang,$^{5, 6}$
	Yingjie Peng,$^{1, 9}$
    Zheng Cai\orcidZ{}$^{10}$
    \\
    $^1$Kavli Institute for Astronomy and Astrophysics, Peking University, Beijing 100871, China\\
	$^2$Department of Astronomy, University of Massachusetts Amherst, MA 01003, USA\\
    $^3$School of Astronomy and Space Science, University of Science and Technology of China, Hefei, Anhui 230026, China\\
    $^4$Key Laboratory for Research in Galaxies and Cosmology, Department of Astronomy, University of Science and Technology of China, Hefei, Anhui 230026, China \\
    $^5$Department of Astronomy, School of Physics and Astronomy, Shanghai Jiao Tong University, Shanghai 200240, China\\
    $^6$Shanghai Key Laboratory for Particle Physics and Cosmology, Shanghai Jiao Tong University, Shanghai 200240, China\\
    $^7$Tsung-Dao Lee Institute, Shanghai Jiao Tong University, Shanghai, 200240, China\\
    $^8$Key Laboratory for Particle Physics, Astrophysics and Cosmology, Ministry of Education, Shanghai Jiao Tong University, Shanghai 200240, China\\
    $^9$Department of Astronomy, School of Physics, Peking University, 5 Yiheyuan Road, Beijing 100871, People’s Republic of China\\
	$^{10}$Department of Astronomy, Tsinghua University, Beijing 100084, China\\
}

\date{Last updated 2020 May 22; in original form 2018 September 5}

\pubyear{2020}

\begin{document}
	\label{firstpage}
	\pagerange{\pageref{firstpage}--\pageref{lastpage}}
	\maketitle


\begin{abstract}
    We propose a novel method to quantify the assembly histories of dark matter
    halos with the redshift evolution of the mass-weighted spatial variance of
    their progenitor halos, i.e. the protohalo size history. We find that the
    protohalo size history for each individual halo at $z\sim 0$ can be
    described by a double power-law function. The amplitude of the fitting
    function strongly correlates to the central-to-total stellar mass ratios of
    descendant halos.  The variation of the amplitude of the protohalo size
    history can induce a strong halo assembly bias effect for massive halos.
    This effect is detectable in observation using the central-to-total stellar
    mass ratio as a proxy of the protohalo size. The correlation to the
    descendant central-to-total stellar mass ratio and the halo assembly bias
    effect seen in the protohalo size are much stronger than that seen in the
    commonly adopted half-mass formation time derived from the mass accretion
    history. This indicates that the information loss caused by the compression
    of halo merger trees to mass accretion histories can be captured by the
    protohalo size history.  Protohalo size thus provides a useful quantity to
    connect protoclusters across cosmic time and to link protoclusters
    with their descendant clusters in observations.
\end{abstract}




\section{Introduction}%
\label{sec:introduction}

In the $\Lambda$CDM cosmological framework, cosmic structures originate from
small primordial perturbations in the very early Universe. These density
fluctuations are then amplified by gravity and eventually form virialized
structures called dark matter halos. The evolution of cosmic structures proceeds
hierarchically, where small halos form first and subsequently
assemble into larger halos \citep{whiteCoreCondensationHeavy1978,
moGalaxyFormationEvolution2010}. In this scenario, gravitational potential
wells associated with dark matter halos provide conditions for the formation of
galaxies, and so galaxies and halos are expected to be tightly connected with
each other. Therefore, understanding the assembly of dark matter halos is a key
step towards understanding the formation and evolution of galaxies
\citep[see][for reviews]{baughPrimerHierarchicalGalaxy2006,
moGalaxyFormationEvolution2010, wechslerConnectionGalaxiesTheir2018}. Dark
matter halos can be characterized from three perspectives.

The first is from the evolution of dark matter halos with cosmic time. The
formation and assembly of dark matter halos can be well-described by halo
merger trees \citep{pressFormationGalaxiesClusters1974,
laceyMergerRatesHierarchical1993, somervilleHowPlantMerger1999}. Such a tree
structure grows from the descendant halo and splits into progenitor halos
backward in time recursively. Halo merger trees are informative but complex. It
is therefore necessary to compress the information by extracting a small set of
important features from full merger trees to describe halo assembly processes.
A common practice is to focus on the main branch by recursively selecting the
main progenitor. The halo mass growth along the main branch is commonly
referred to as the mass accretion history
\citep{wechslerConcentrationsDarkHalos2002,zhaoACCURATEUNIVERSALMODELS2009,
katsianisModellingMassAccretion2023}. A characteristic halo formation time can
be defined as the highest redshift when the halo has accreted half of its final
halo mass, or the redshift when the halo switches from fast accretion to slow
accretion \citep{wechslerConcentrationsDarkHalos2002,
    zhaoGrowthStructureDark2003, gaoAgeDependenceHalo2005,
wechslerDependenceHaloClustering2006}.

The second perspective on dark matter halos is their internal structure.
\citet{navarroUniversalDensityProfile1997} found that the density profile of
dark matter halos can be well-described by a universal profile
\begin{equation}
    \rho(r) = \frac{\rho_s}{(r/r_s)(1 + r/r_s)^2}
\end{equation}
where $\rho_s$ is the density parameter and $r_s$ is a scaling radius. The
asymptotic behavior of this profile is $\rho\propto r^{-1}$ for $r\ll r_s$ and
$\rho \propto r^{-3}$ for $r\gg r_s$. In practice, a dark matter halo is
restrained within a radius, $R_{\rm vir}$, within which the mean density is
equal to some chosen value, and the total mass within $R_{\rm vir}$ is defined
as the halo mass, $M_{\rm vir}$. Therefore, the NFW halo can be characterized
with two other variables: the halo mass $M_{\rm vir}$ and the halo
concentration $c=R_{\rm vir}/r_s$. In addition, dark matter halos also exhibit
non-spherical shape and the deviation from the spherical symmetry is more
prominent for massive halos \citep{jingTriaxialModelingHalo2002,
allgoodShapeDarkMatter2006}. Finally, dark matter halos also possess
substructures from undissolved recent halo mergers
\citep{mooreDarkMatterSubstructure1999, yangANALYTICALMODELACCRETION2011,
jiangStatisticsDarkMatter2016}.

The last perspective on the halo population is from the spatial distribution,
or the clustering of halos. The spatial distribution of dark matter halos can
be characterized by the two-point correlation function and higher-order
statistics. The primary feature of halo clustering is the dependence on the
halo mass, where massive halos are more clustered than low-mass halos. This
phenomenon can be explained with a Gaussian initial density field and the
extended Press-Schechter formalism \citep{moAnalyticModelSpatial1996,
shethEllipsoidalCollapseImproved2001}. Moreover, the clustering of dark matter
halos also depends on their secondary properties, which is usually referred to
as the halo assembly bias. \citet{gaoAgeDependenceHalo2005} first identified
the dependence on halo formation time \citep[see
also][]{wechslerDependenceHaloClustering2006, liHaloFormationTimes2008,
    wangDistributionEjectedSubhaloes2009, mansfieldThreeCausesLowmass2020,
wangEvaluatingOriginsSecondary2021a, cite1, cite2, cite3}. Subsequent studies
also find halo assembly bias caused by halo concentration, halo spin, and other
properties \citep[e.g.][]{gaoAssemblyBiasClustering2007,
jingDependenceDarkHalo2007, wangInternalPropertiesEnvironments2011}. There are
also attempts to detect such effects in observations, with mixed results
\citep[e.g.][]{wangDetectionGalaxyAssembly2013, zuDoesConcentrationDrive2021,
wangEvidenceGalaxyAssembly2022}.

A comprehensive understanding of dark matter halos requires not only knowledges
from each perspective, but also relationships among properties in these three
categories. In this respect, one important question is how halo structures are
shaped by their assembly histories. \citet{navarroUniversalDensityProfile1997}
proposes a simple model that the difference in halo formation time and the
time-dependence of cosmic density result in different halo concentrations. This
model is further improved by subsequent studies
\citep{bullockProfilesDarkHaloes2001, wechslerConcentrationsDarkHalos2002,
    zhaoGrowthStructureDark2003, zhaoMassRedshiftDependence2003,
    luOriginColdDark2006, zhaoACCURATEUNIVERSALMODELS2009,
diemerAccuratePhysicalModel2019}. Recently,
\citet{wangConcentrationsDarkHaloes2020} investigated the importance of
merger events in shaping the halo concentration, and found that secular
evolutions increase the concentration while sudden halo mergers reduce the
concentration.

There is, however, one unresolved question on the relationship among halo
structure, halo assembly history, and halo clustering.
\citet{jingDependenceDarkHalo2007} found that the halo assembly bias induced by
the halo concentration is much stronger than the halo formation time for halos
above $\gtrsim 10^{13}h^{-1}\rm M_{\odot}$. A detailed analysis in
\citet{maoAssemblyBiasExploring2018} revealed that paired and unpaired
cluster-size halos have nearly identical mass accretion histories, and so there
is no correlation between halo clustering and halo formation time, in contrast
to results obtained by \cite{chueAssemblyRequiredAssembly2018}. The question
is: if the halo structure is determined by its assembly history, why the
assembly bias of the halo concentration is absent in the mass accretion history
for cluster-size halos? One possible reason is that the assembly history alone
cannot determine the structure of descendant halos
\citep[see][]{ludlowDynamicalStateMassconcentration2012}. Another explanation
is that the information about the halo assembly bias is lost during the data
compression from halo merger trees to mass accretion histories.
\citet{wangEvaluatingOriginsSecondary2021a} proposed that the halo formation
time is determined by both internal and external factors, which can produce
opposite halo assembly bias effects with similar amplitudes for massive halos.
Therefore, the halo assembly bias effect of the halo formation time is very
weak. In that case, a different compression method might be needed to retain
the information about halo assembly bias.

In this study, we propose a new method to characterize the assembly of dark
matter halos, using the redshift evolution of protohalo sizes. This paper is
organized as follows. \S\,\ref{sec:data} introduces the simulation data used in
this study.
\S\,\ref{sec:protohalo_size_evolution_and_its_relation_to_descendant_halos}
presents the redshift evolution of protohalo sizes and their relation to
descendant halos. \S\,\ref{sec:halo_assembly_bias} shows the assembly bias
effect in terms of protohalo size. Finally, the discussion and summary of our
main results are presented in \S\,\ref{sec:discussion_and_summary}.

\section{Data}%
\label{sec:data}

In this study, we use the ELUCID simulation
\citep{wangRECONSTRUCTINGINITIALDENSITY2013, wangELUCIDEXPLORINGLOCAL2014,
wangELUCIDEXPLORINGLOCAL2016, tweedELUCIDExploringLocal2017}, which is a
constrained N-body simulation to reconstruct the density field and formation
history of our local Universe based on the Sloan Digital Sky Survey DR7
\citep{yorkSloanDigitalSky2000, abazajianSeventhDataRelease2009}. This
simulation has $3072^3$ dark matter particles, each  with a mass of $3.09\times
10^8h^{-1}\rm M_\odot$, in a box with a side length of $500h^{-1}\rm Mpc$. The
simulation assumes a $\Lambda$CDM cosmology with $\Omega_m = 0.258$,
$\Omega_\Lambda=0.742$, $\sigma_8=0.80$, and $h = 0.72$.

Dark matter halos are identified using the Friends-of-Friends algorithm
\citep{davisEvolutionLargescaleStructure1985}, and their masses are assigned as
the total dark matter mass enclosed within a radius where the mean overdensity
is 200 times of the critical density. The ELUCID simulation is complete for
halos to $10^{10}h^{-1}\rm M_\odot$ up to $z\sim 8$
\citep{wangELUCIDEXPLORINGLOCAL2016}. The concentration, $c= R_{\rm vir}/r_s$,
of each FoF halo is estimated through their first moment, i.e.
$R_1=\int_0^{R_{\rm vir}}4\pi r^3\rho(r)dr/M_{\rm vir}/R_{\rm vir}$, whose
relation to the concentration of NFW halos can be expressed analytically
\citep[see][for details]{wangconcnetration2023}. In each FoF halo, subhalos are
identified with the SUBFIND algorithm
\citep{springelPopulatingClusterGalaxies2001}, where the most massive subhalo
is defined as the central subhalo and the remaining ones are satellite
subhalos. Subhalos are linked to their progenitors and descendants using the
code provided by \citet{springelSimulationsFormationEvolution2005}. The main
branch of each subhalo is defined by recursively selecting the most massive
progenitor. The peak halo mass, denoted as $M_{\rm peak}$, for each subhalo is
defined as the maximum halo mass that it has achieved when it is a central
subhalo. Finally, a protohalo is defined as the collection of all progenitor
halos at a given $z> 0$ that would end up in a common descendant halo at $z=0$
\citep{wangFindingProtoclustersTrace2021}. Stellar masses are assigned to
individual subhalos following the stellar mass-halo mass relation in
\texttt{UniverseMachine} based on $M_{\rm peak}$ for each subhalo \citep[see
Eq.~J1 in][]{behrooziUNIVERSEMACHINECorrelationGalaxy2019}. For each cluster at
$z=0$, we denoted its central stellar mass as $M_{*, \rm cen}$, and its total
stellar mass as $M_{*, \rm tot}$. In this study, we use all dark matter halos
above $10^{13}h^{-1}\rm M_\odot$ at $z=0$ and trace their progenitor halos
above $10^{10}h^{-1}\rm M_\odot$ to $z=8$.

\section{The protohalo size history and its relation to descendant halos}%
\label{sec:protohalo_size_evolution_and_its_relation_to_descendant_halos}

\subsection{Protohalo size and its redshift evolution}%
\label{sub:protohalo_size_and_its_redshift_evolution}

\begin{figure*}
    \centering
    \includegraphics[width=1\linewidth]{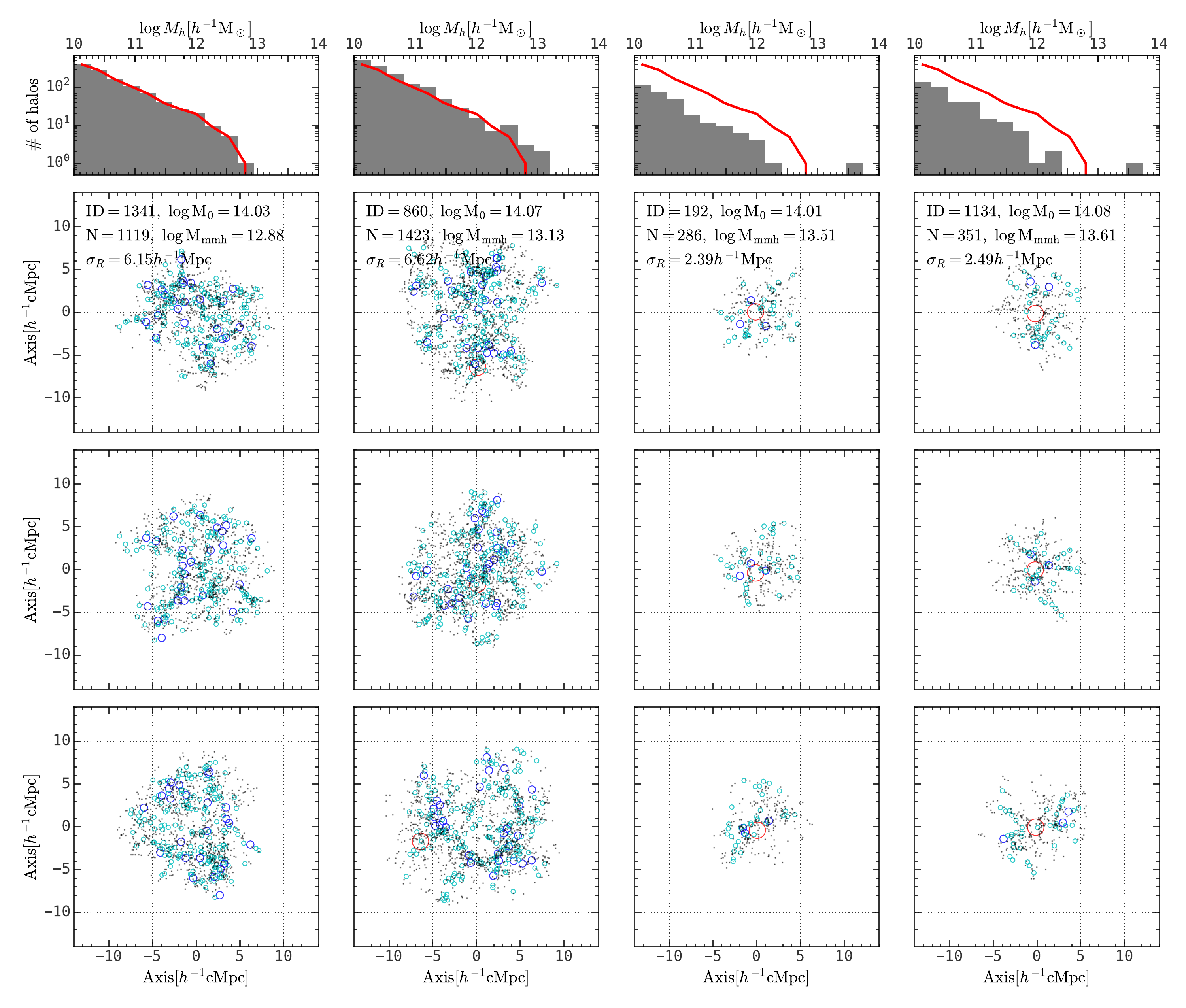}
    \caption{
        The member halo spatial distribution for four example protohalos at
        $z\sim 2$ with descendant halo mass around $10^{14}h^{-1}\rm M_\odot$.
        The left two columns are two examples with large protohalo sizes, and
        the right two columns are two examples with small protohalo sizes. The
        top panels show the halo mass distribution in each protohalo, and the
        halo mass distribution of the leftmost protohalo is overplotted on each
        panel in red solid curves for reference. The bottom three rows are for
        different projection directions. Halos above $10^{13}$, $10^{12}$, and
        $10^{11}h^{-1}\rm M_\odot$ are shown in red, blue, and cyan circles,
        and halos within $10^{10}-10^{11}h^{-1}\rm M_\odot$ are shown in black
        dots. The information about each halo, including the ID in ELUCID, the
        descendant halo mass, the richness, the halo mass of the most massive
        halo, the protohalo size, is shown in each row. This figure
        demonstrates the large diversity of spatial distribution of protohalos
        at given descendant halo mass.
    }%
    \label{fig:figures/examples_of_pc_diff_z2}
\end{figure*}

\begin{figure*}
    \centering
    \includegraphics[width=1\linewidth]{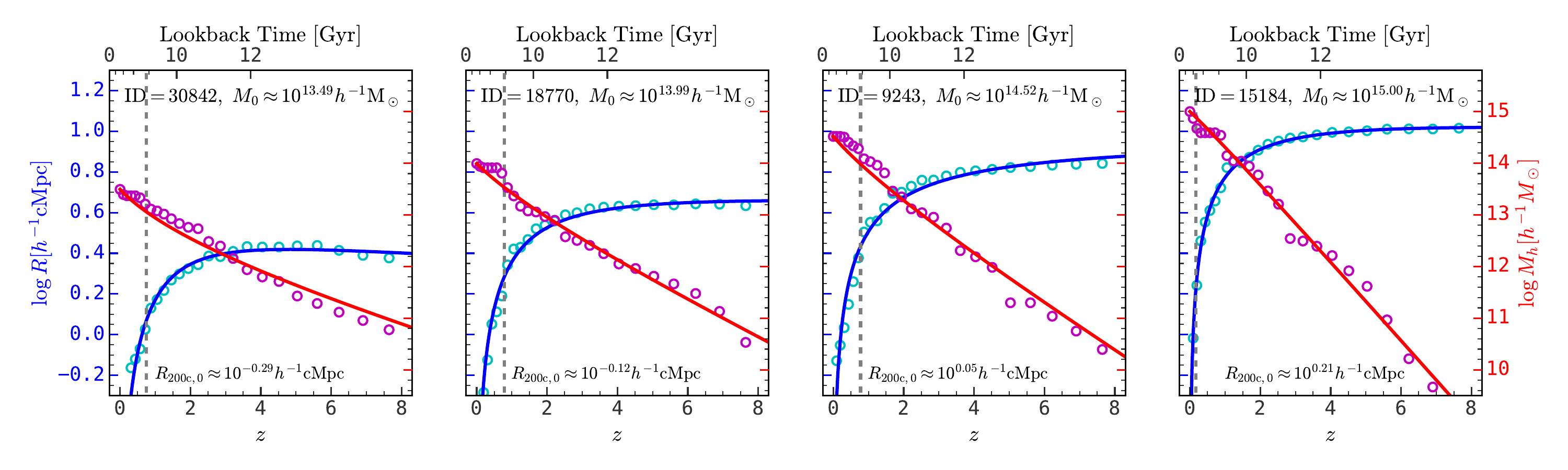}
    \caption{
        The protohalo size histories (cyan circles and blue solid lines on the
        left $y$-axis) and the mass accretion histories (magenta circles
        and red solid lines on the right $y$-axis) for four halos
        selected at $z\sim 0$. The circles are the histories calculated from
        each snapshot, and the solid lines are the fitting functions from
        equations~\ref{eq:ph_size_evolution} and \ref{eq:mah}. The vertical
        dashed lines show the half-mass time $z_{\rm half}$ for each halo.
    }%
    \label{fig:figures/size_evolution_mah_examples}
\end{figure*}

\begin{figure*}
    \centering
    \includegraphics[width=1\linewidth]{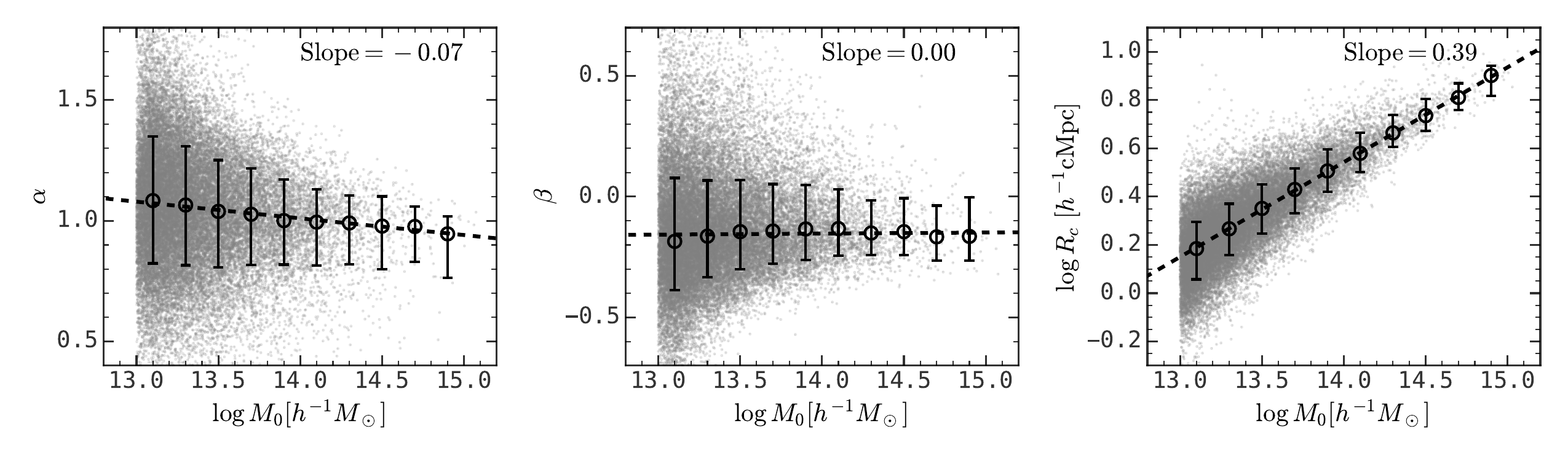}
    \caption{
        The scatter of the descendant halo mass and three parameters for the
        protohalo size histories in equation~\ref{eq:ph_size_evolution}. The
        error bars show the median and the $16^{\rm th}-84^{\rm th}$
        percentiles. The black dashed lines are the linear fitting
        functions to the median values, where the slope is labeled on each
        panel.
    }%
    \label{fig:figures/protohalo_size_depend_m0}
\end{figure*}

The assembly of a dark matter halo can be fully characterized by its merger
tree. A snapshot of a halo merger tree at $z> 0$ gives a collection of halos
that will eventually end up in the halo at $z=0$. Here we define this
collection of halos at $z>0$ as a protohalo of its descendant halo at $z=0$
\citep{wangFindingProtoclustersTrace2021}. We define the size of a protohalo as
\begin{equation}
    R = {\sqrt{\sum m_i\|\mathbf x_i - \mathbf x_{\rm cen}\|^2\over \sum m_i}},
    ~~~\mathbf x_{\rm cen} \equiv {\sum_im_i\mathbf x_i\over \sum_i m_i}
    \label{eq:ph_size}
\end{equation}
where $m_i$ and $\mathbf x_i$ are the mass and comoving position of the $i$-th
progenitor halo, $\mathbf x_{\rm cen}$ is the center of mass, and the sum is
for all halos above some mass threshold. We trace all the progenitor subhalos
for each descendant subhalo contained in the $z=0$ main halo, from which we
select all central subhalos at a given redshift and use their $M_{\rm 200c}$ as
$m_i$ to perform the summation in equation~(\ref{eq:ph_size}). For our main
presentation, we choose the threshold to be $10^{10}h^{-1}\rm M_\odot$ on the
basis of ELUCID resolution. In
Appendix~\ref{sec:protohalo_sizes_with_different_halo_mass_limits}, we show
that the ranking of protohalo sizes is nearly unaffected by the choice of the
mass threshold. In addition, we also show in Appendix~\ref{sec:centering} that
using the center determined by a few dominating halos to replace $\mathbf
x_{\rm cen}$ does not alter the ranking of protohalo sizes significantly.

Fig.~\ref{fig:figures/examples_of_pc_diff_z2} shows the spatial distribution of
member halos for four protohalos with similar descendant halo mass but with
different sizes at $z\sim 2$. The left and right two columns are two examples
of protohalos with large and small sizes, respectively. The three rows show
projections in three directions, and the origins are chosen to be the center of
mass, $\mathbf x_{\rm cen}$. Here the red, blue, and cyan circles represent
halos with mass above $10^{13}$, $10^{12}$ and $10^{11}h^{-1}\rm M_\odot$,
respectively, and black dots are for halos in the mass range of
$10^{10}-10^{11}h^{-1}\rm M_\odot$. Despite the fact that these protohalos will
eventually collapse to form descendant halos of similar masses at $z=0$, their
spatial distributions at high $z$ are very diverse: the difference in linear
size between the two extremes could be as large as a factor of $\sim 3$.
Large protohalos (the left two columns) show diverse morphologies with abundant
low-mass halos, and their most massive halos are not so different from other
massive member halos. In contrast, small protohalos (the right two columns) are
more compact with less abundant member halos. They usually have a dominating
halo, whose mass can be 10 times higher than the second one, located near the
center of mass. The diversity in protohalo sizes motivates us to investigate
the information on halo assembly and structure encapsulated in this quantity.

Fig.~\ref{fig:figures/size_evolution_mah_examples} shows the redshift evolution
of the protohalo size of four halos selected at $z\sim 0$ with different halo
masses. First of all, the redshift evolution of protohalo size is well-behaved.
It is nearly a constant above $z\sim 2$, below which it rapidly collapses into
a virialized halo by $z=0$. Most strikingly, the turnover point occurs at
$z\sim 2$ across the whole halo mass range (see also
Fig.~\ref{fig:figures/protohalo_size_evolution}). Secondly, the protohalo size
above $z\sim 2$ monotonically increases with the descendant halo mass. And the
protohalo size is $\sim 2\rm Mpc$ for group-size halos, $\sim 5\rm Mpc$ for
``Fornax''-like halos, $\sim 6\rm Mpc$ for ``Virgo''-like halos, and $\gtrsim
10\rm Mpc$ for ``Coma''-like halos. The correlation between protohalo size and
the descendant halo mass is expected, because the formation of a more massive halo
requires matter distributed over a larger volume at high $z$ due to the
homogeneity of the early Universe. Finally, the redshift evolution of protohalo
size is rather smooth.

Fig.~\ref{fig:figures/size_evolution_mah_examples} also shows the mass
accretion history, which is the halo mass evolution on the main branch, in
magenta circles. The halo mass difference between current halos and their main
progenitors at $z\sim 8$ is up to 3-5 dex, indicating that the mass accretion
history based on the main branch only captures a small portion of the whole
protohalo at high $z$. The dashed vertical line in each panel shows the
half-mass redshift when the main progenitor has achieved half of its final halo mass.
The small values of the half-mass redshift indicate that they only characterize the
late-time evolution of halos.

The protohalo size history and the mass accretion history are two ways to
compress the halo merger tree into a one-dimensional function, and further
compression may be achieved by fitting these histories with
some simple parametric forms. Motivated by the two-stage evolution of
protohalo size observed above, we use a double power-law function to fit its
evolution from $z\sim 8$ to $z\sim 0$:
\begin{equation}
    R(z) = {2R_c (z/z_c)^\alpha \over 1 + (z/z_c)^{\alpha-\beta}}
    \label{eq:ph_size_evolution}
\end{equation}
where $R_c = R(z_c)$ is the amplitude of the protohalo size history.  If
$\alpha> \beta$, then $R(z)\propto z^\alpha$ for $z\ll z_c$ and $R(z)\propto
z^\beta$ for $z\gg z_c$. We assume prior ranges of $\alpha\in (0, 5)$ and
$\beta \in (-1, 1)$ for $\alpha$ and $\beta$, respectively, and set $z_c=2$ as
it is sufficient to fit the data (see
Fig.~\ref{fig:figures/size_evolution_mah_examples}). This functional form can
accurately describe individual protohalo size history across the whole halo
mass range probed according to our visual inspection.

Fig.~\ref{fig:figures/protohalo_size_depend_m0} shows the dependence of fitting
parameters on the descendant halo mass. Here one can see that the late-time
slope $\alpha$ is around 1 and it has a marginal dependence on the descendant
halo mass, with low-mass halos tending to collapse slightly more rapidly. The
early-time slope $\beta$ is independent of the descendant halo mass, and has a
median value of $-0.2$. Finally, the amplitude $R_c$ has a power-law relation
with the descendant halo mass, with a logarithmic slope of $\sim 0.39$. In an ideal
situation where descendant halos form from a uniform density field, we expect
${\rm d}\log R_c/{\rm d}\log M_0 = 1/3\approx 0.33$, which is close to our
result here.

\subsection{Relation to the structure of descendant halos}%
\label{sub:relation_to_descendant_halos}

\begin{figure*}
    \centering
    \includegraphics[width=1\linewidth]{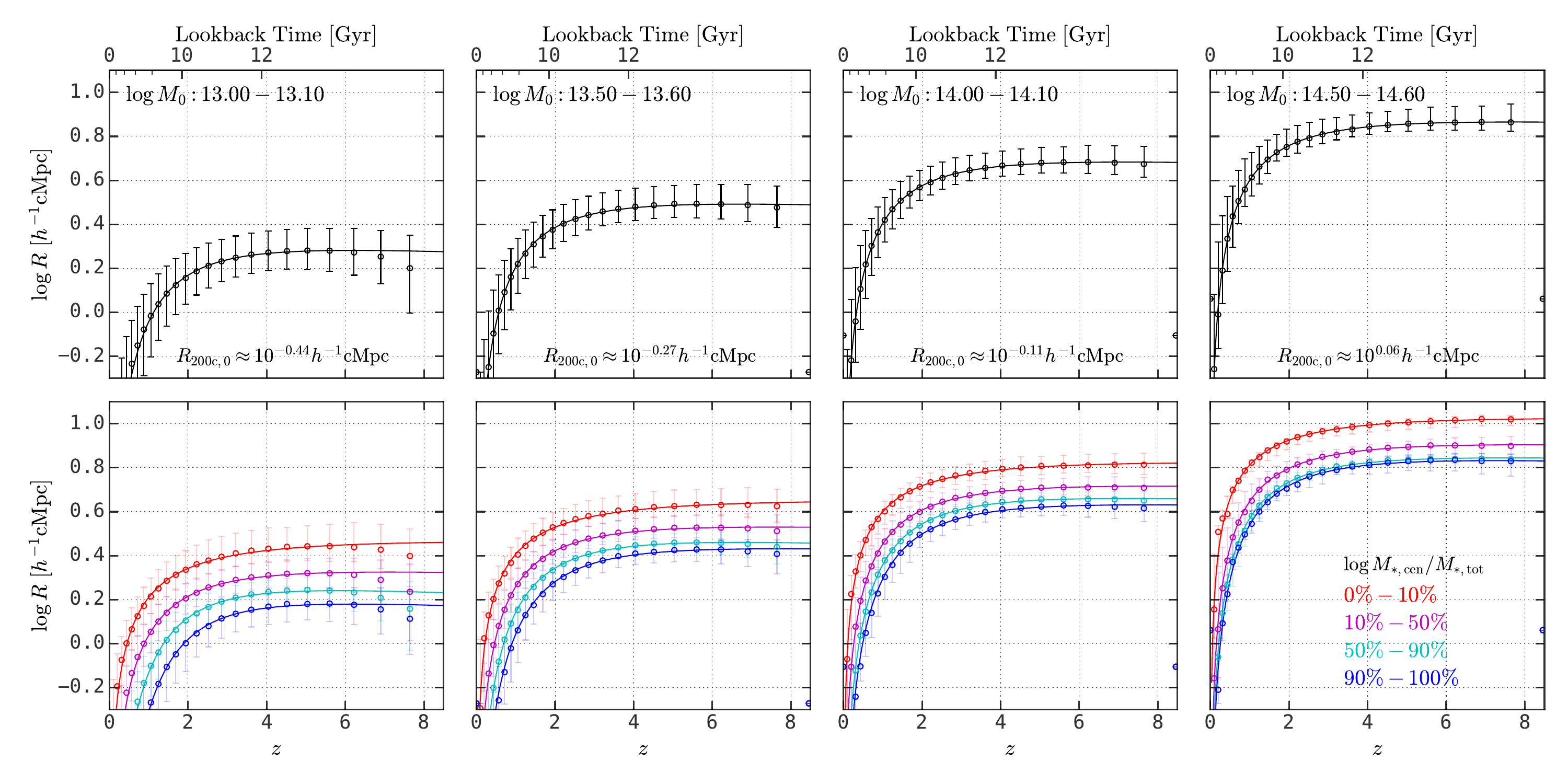}
    \caption{
        Top panels: The redshift evolution of protohalo size, $R$, in
        four halo mass bins. Bottom panels: In each halo mass bin, halos
        are divided into four equal-size subsamples according to their $\log
        M_{*, \rm cen}/M_{*, \rm tot}$. The error bars show the median and the
        $16^{\rm th}-84^{\rm th}$ percentiles. This figure demonstrates the
        strong correlation between the protohalo size history and the $\log
        M_{*, \rm cen}/M_{*, \rm tot}$ of descendant halos.
    }%
    \label{fig:figures/protohalo_size_evolution}
\end{figure*}

\begin{figure*}
    \centering
    \includegraphics[width=1\linewidth]{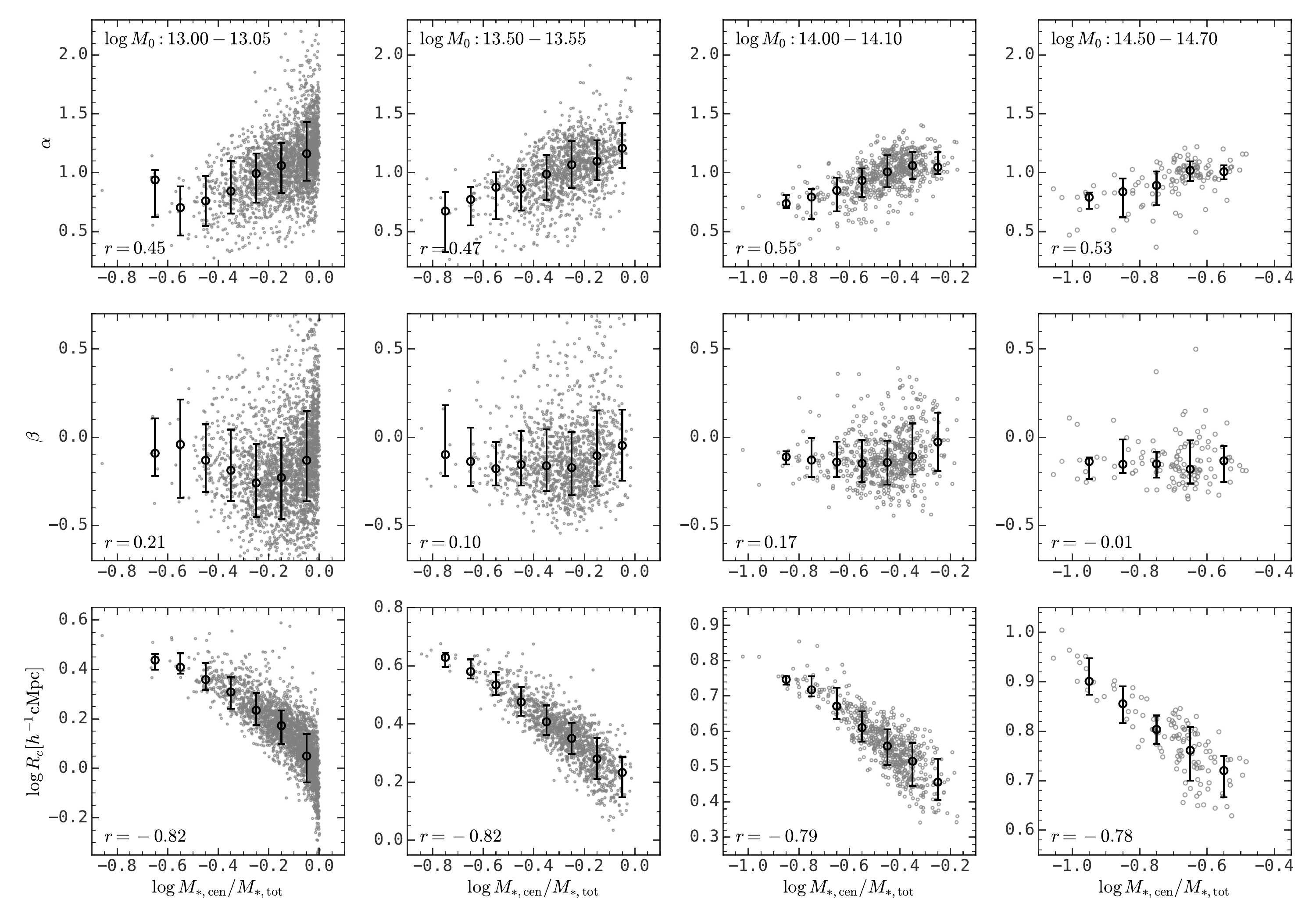}
    \caption{
        The correlation between the central-to-total stellar mass ratio, $\log
        M_{*, \rm cen}/M_{*, \rm tot}$, and the parameters in
        equation~\ref{eq:ph_size_evolution}. Four columns are for different
        descendant halo mass bins, and the bin sizes are adjusted to be
        minimized while containing sufficient data points for statistical
        analysis. The top panels are for the late-time slope $\alpha$,
        the middle panels are for the early-time slope $\beta$, and the
        bottom panels are for the amplitude $\log R_c$. This figure
        demonstrates that the amplitude of the protohalo size history strongly
        correlates to the descendant central-to-total stellar mass ratio, as
        the correlation for the other two parameters is much weaker.
    }%
    \label{fig:figures/protohalo_size_correlation_in_mass_bin}
\end{figure*}

\begin{figure}
    \centering
    \includegraphics[width=0.9\linewidth]{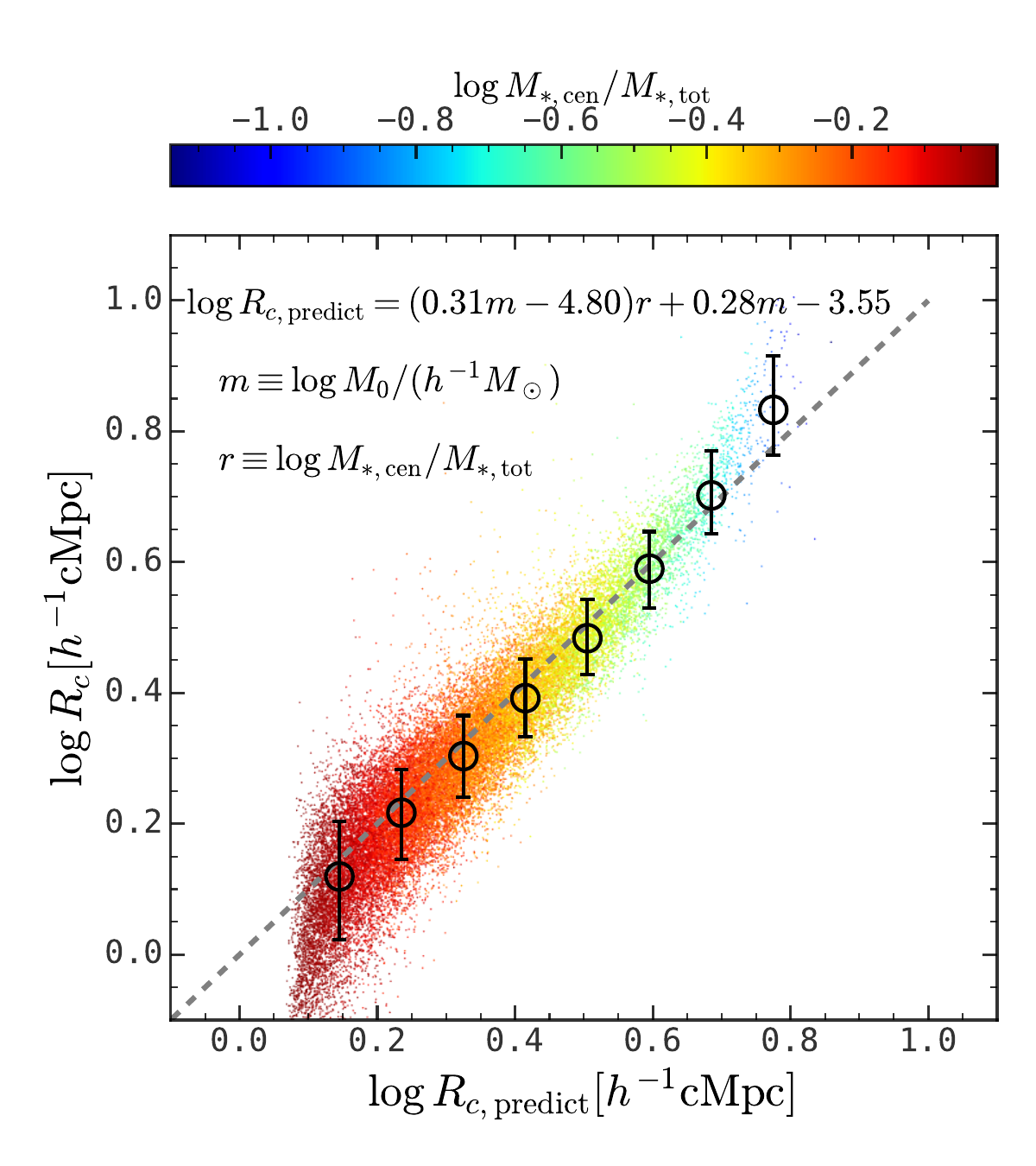}
    \caption{
        Accuracy of the amplitude of the protohalo size history, i.e. $\log
        R_c$, calibrated in equation~\ref{eq:calibrate_rc}. The $x$-axis is the
        prediction from the relation shown in the figure, and the $y$-axis is
        the true $\log R_c$ for each individual halo. The color encodes the
        central-to-total stellar mass ratio. The error bars show the median and
        the $16^{\rm th}-84^{\rm th}$ percentiles. This figure shows that the
        $\log R_c$ calibrated against $\log M_0$ and $\log M_{*, \rm cen}/M_{*,
        \rm tot}$ matches the true $\log R_c$ quite well.
    }%
    \label{fig:figures/predict_rc}
\end{figure}

\begin{figure}
    \centering
    \includegraphics[width=0.9\linewidth]{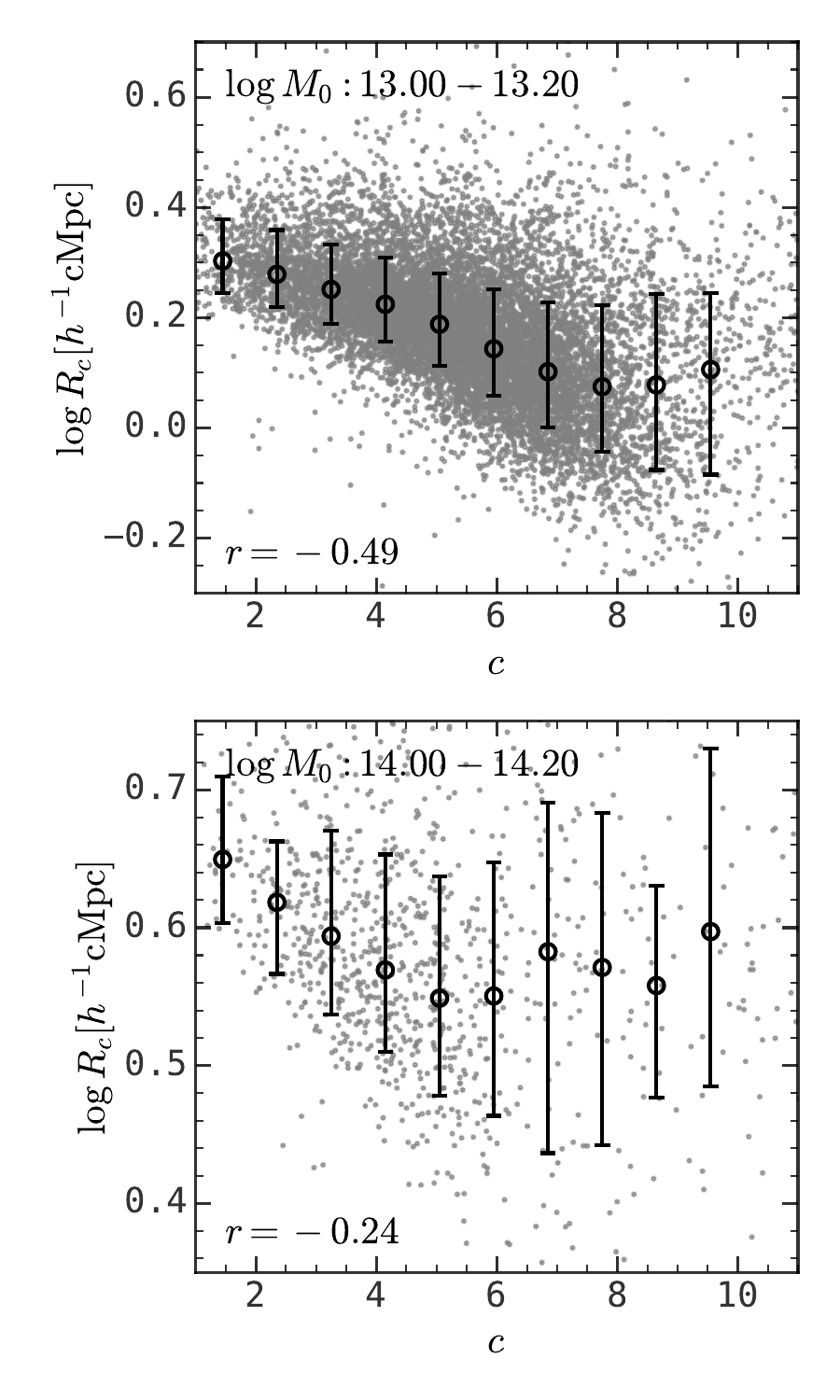}
    \caption{
        Relation between the amplitude of the protohalo size history, $\log
        R_c$, and the halo concentration, $c$, in two halo mass bins. The
        erorr bars show the median and the $16^{\rm th}-84^{\rm th}$
        percentiles. This figure shows that $\log R_c$ negatively correlates
        to $c$.
    }%
    \label{fig:figures/correlation_loga_conc}
\end{figure}

Here we first study the relation between the protohalo size history and the
substructure of descendant halos. We adopt the central-to-total stellar mass ratio
to characterize the substructure, for the following reasons. First of all,
this quantity is observable, so that the results obtained here can be related to
observations. Secondly, the stellar mass used here is obtained according to the
$M_{\rm peak}$ of each subhalo (see \S\,\ref{sec:data}) and this relation is
well-established at $z\sim 0$ \citep{wechslerConnectionGalaxiesTheir2018}. Finally,
$M_{\rm peak}$ is less affected by the resolution of simulations, while other
quantities for substructures, such as the subhalo mass, require high resolution
for reliable estimates \citep{jiangStatisticsDarkMatter2016}.

We find that the protohalo size history strongly correlates with the
substructures of descendant halos.
Fig.~\ref{fig:figures/protohalo_size_evolution} shows the average protohalo
size evolution as a function of redshift with different descendant halo masses,
$\log M_0$, where the error bars show the median and the $16^{\rm th}-84^{\rm
th}$ percentiles and the solid lines are the fitting results of the double
power-law function. In each bottom panel, descendant halos are further divided
into four subsamples according to their central-to-total stellar mass ratios,
$\log M_{*, \rm cen}/M_{*, \rm tot}$, where the stellar mass is assigned using
the empirical relations in \texttt{UniverseMachine}. At $z> 2$, the amplitude
of the protohalo size history has a strong dependence on the central-to-total
stellar mass ratio of descendant halos, where halos with higher $\log M_{*, \rm
cen}/M_{*, \rm tot}$ tend to have smaller protohalos. Quantitatively, the
difference in the protohalo size for halos with the top-10\% and bottom-10\%
central-to-total stellar mass ratios is about 0.2-0.3 dex, despite that the
$1-\sigma$ range of the protohalo size in each descendant halo mass bin is only
$\sim$0.1-0.2 dex (see top panels).

A quantitative correlation analysis is presented in
Fig.~\ref{fig:figures/protohalo_size_correlation_in_mass_bin}, which shows the
relation between the parameters in equation~\ref{eq:ph_size_evolution} and the
central-to-total stellar mass ratio, $\log M_{*, \rm cen}/M_{*, \rm tot}$, of
descendant halos in four narrow halo mass bins. The top panels show the results
for the late-time slope $\alpha$, where Spearman's rank correlation coefficient
is about 0.5 across the whole halo mass range. Descendant halos with more
dominating central galaxies tend to have steeper late-time slopes,
indicating more rapid collapses of protohalos to form descendant halos. The panels
in the middle row show results for the early-time slope $\beta$, whose
correlation with $\log M_{*, \rm cen}/M_{*, \rm tot}$ is negligible. Finally, the
bottom panels show results for the amplitude of the protohalo size history,
$\log R_c$, where Spearman's rank correlation coefficient is about 0.8. The
difference in $\log R_c$ for halos with the highest and lowest $\log M_{*, \rm
cen}/M_{*, \rm tot}$ is about 0.4 dex, compared to the $1-\sigma$ width of
$\log R_c$ in each halo mass bin, which is about 0.1-0.2 dex (see the top
panels of Fig.~\ref{fig:figures/protohalo_size_evolution}). In addition, we
find that the dependence of $\log R_c$ on $\log M_0$ and $\log M_{*, \rm
cen}/M_{*, \rm tot}$ can be described concisely by
\begin{align} \label{eq:calibrate_rc} \log\frac{R_c}{h^{-1}{\rm Mpc}} =
    a_1\times\log\frac{M_{*, \rm cen}}{M_{*, \rm tot}} + a_2\\ a_1 =
    0.31\times\log\frac{M_0}{h^{-1}\rm M_\odot} - 4.80 \nonumber\\
    a_2 = 0.28\times\log\frac{M_0}{h^{-1}\rm M_\odot} - 3.55 \nonumber
\end{align}
As one can see from Fig.~\ref{fig:figures/predict_rc}, the predicted $\log R_c$
matches the true $\log R_c$ well and the width for the $16^{\rm
th}-84^{\rm th}$ percentiles is smaller than 0.1 dex.

In Appendix~\ref{sec:protohalo_sizes_using_different_subhalo_definition}, we
restrained the protohalo as the collection of halos that will end up in
descendant subhalos within the virial radius of the descendant main halo. This
will effectively eliminate member halos on the protohalo outskirt, since they
are likely to become subhalso outside the descendants' virial radius. We found
that the correlation between the protohalo size and the central-to-total
stellar mass ratio is compromised but still as high as $\approx 0.6$.

In Appendix~\ref{sec:results_for_the_illustrstng_simulation}, we perform a
similar analysis using the IllustrisTNG simulation \citep{Pillepich_2018a} and
obtain similar results to Figs.~\ref{fig:figures/protohalo_size_evolution} and
\ref{fig:figures/protohalo_size_correlation_in_mass_bin}, which indicates that
the results obtained here do not depend on the specific stellar mass-halo
mass relation in \texttt{UniverseMachine}.

The relation between $\log R_c$ and $\log M_{*, \rm cen}/M_{*, \rm tot}$ can be
explained qualitatively. To begin with, the total mass in the region occupied
by a protohalo is proportional to its final descendant halo mass. At a
given descendant halo mass, protohalos with larger sizes collapse at later
times due to their relatively shallow gravitational potential. Therefore, the
halo-halo merger events occur relatively late, and it becomes less probable for
these late-accreted satellite subhalos and their galaxies to be cannibalized by
their central subhalos and central galaxies. Consequently, these systems have
more substructures, which results in low values of $\log M_{*,
\rm cen}/M_{*, \rm tot}$.

Finally, Fig.~\ref{fig:figures/correlation_loga_conc} shows a moderate negative
correlation between $\log R_c$ and the halo concentration, $c$, where
high-concentration halos have more compact protohalos. The rank correlation
coefficient is $\gtrsim 0.5$ for group-size halos and decreases to 0.3 for
cluster-size halos. This result is consistent with the result in
\citet{wangEvaluatingOriginsSecondary2021a}, where it was found that
high-concentration halos have higher overdensity in Lagrangian space.

\subsection{Relation to the mass accretion history}%
\label{sub:relation_to_the_mass_accretion_history}

\begin{figure}
    \centering
    \includegraphics[width=1\linewidth]{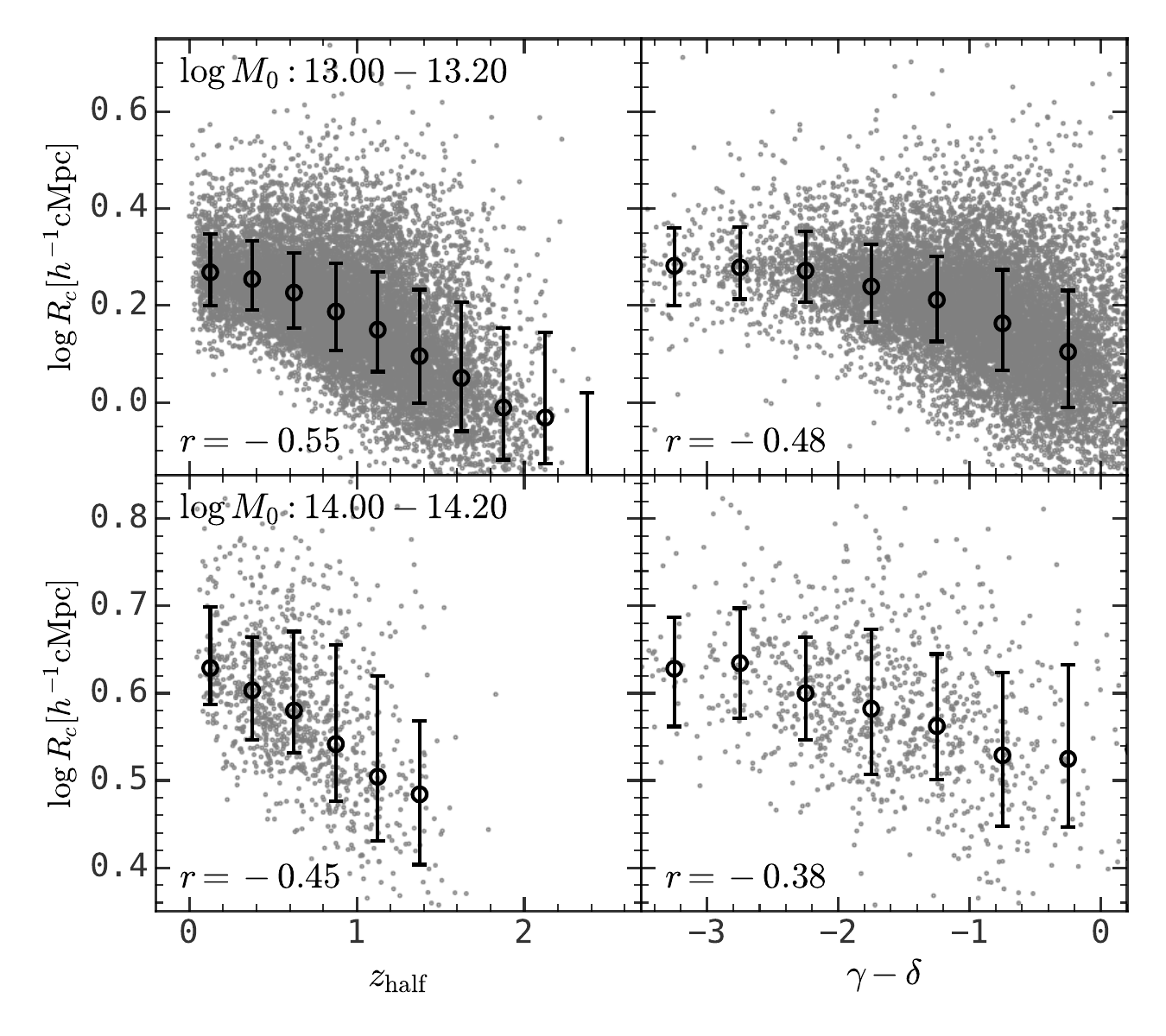}
    \caption{
        The relation between the amplitude of the protohalo size history, $\log
        R_c$, and characteristic features of massive accretion history,
        including the half-mass time, $z_{\rm half}$, and the recent halo
        accretion rate, $\gamma - \delta$, (see equation~\ref{eq:acc_rate}). The
        error bars show the median and $16^{\rm th}-84^{\rm th}$ percentiles.
        This figure shows that the halos with large protohalo sizes tend to form
        later and experience more rapid accretion than the halos with small
        protohalo size.
    }%
    \label{fig:figures/correlation_loga_mah}
\end{figure}

\begin{figure}
    \centering
    \includegraphics[width=0.9\linewidth]{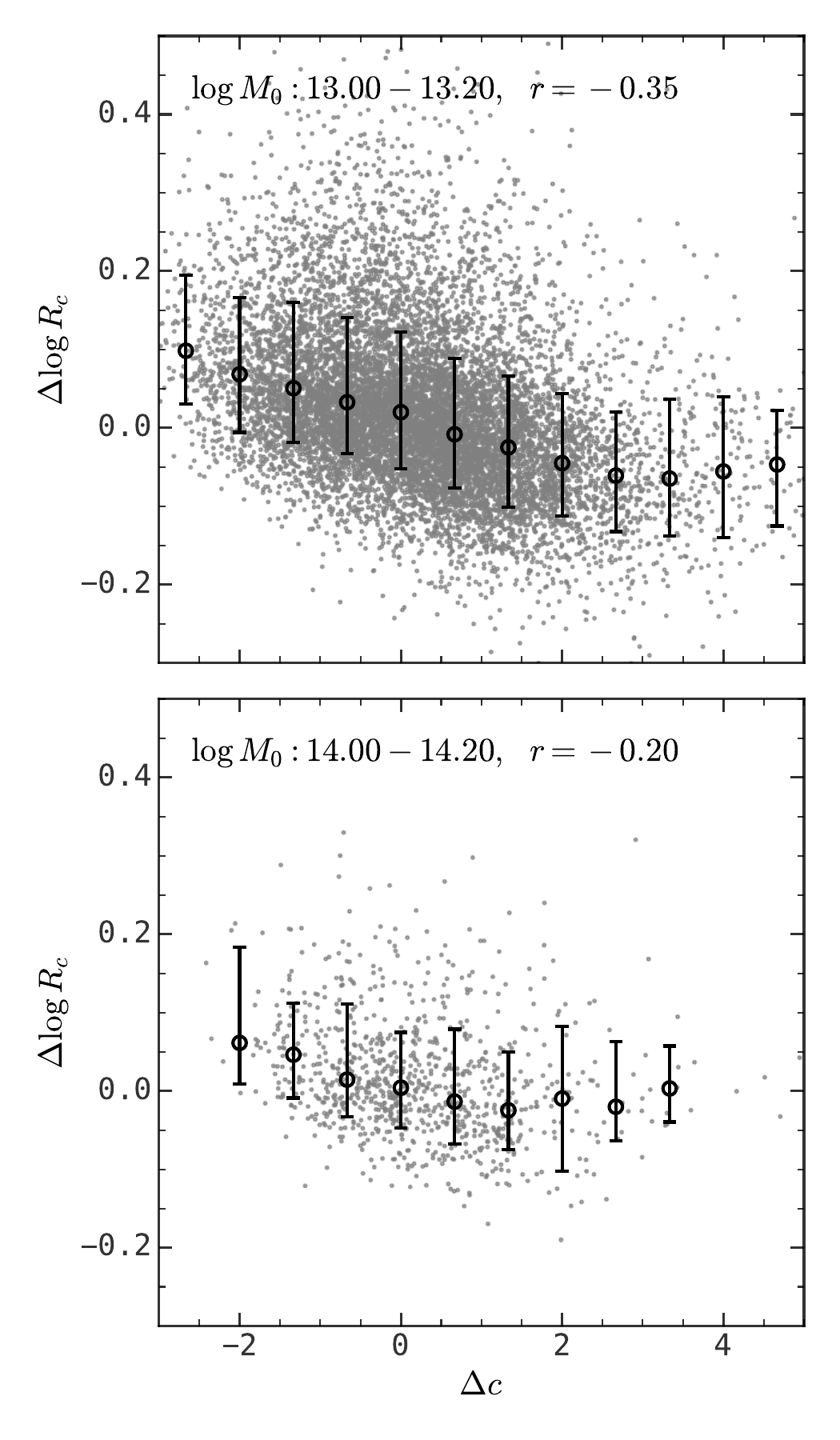}
    \caption{
        The correlation between $\Delta c$ and $\Delta\log R_c$ in two narrow
        halo mass bins. Here $\Delta c$ and $\Delta\log R_c$ are the difference
        between these two quantities and their median relation with respect to
        $z_{\rm half}$. The error bars show the median and $16^{\rm th}-84^{\rm
        th}$ percentiles. This figure demonstrates that the amplitude of the
        protohalo size history contains extra information about the descendant
        halo structure that is missed by the half-mass time.
    }%
    \label{fig:figures/correlation_zhalf_ph_size_conc2}
\end{figure}

The hierarchical formation process of dark matter halos can be fully
characterized by halo merger trees. However, these tree-like structures are too
complicated to use directly in the modeling of halo assembly. Some data
compression is required. There have been attempts to extract compressed
information from halo merger trees. A common method is to compress the halo
merger tree into the mass accretion history
\citep{wechslerConcentrationsDarkHalos2002, zhaoGrowthStructureDark2003}.
Further data compression can be done by defining some characteristic
formation time, fitting the mass accretion history with some simple parametric
forms, or using the principle component analysis technique
\citep{wechslerConcentrationsDarkHalos2002, gaoAgeDependenceHalo2005,
    wechslerDependenceHaloClustering2006, liHaloFormationTimes2008,
chenRelatingStructureDark2020}. Here we fit the mass accretion history with the
functional form in \citet{mcbrideMassAccretionRates2009}, which is
\begin{equation}
    M(z) = M_0(1 + z)^\gamma e^{-\delta z} \label{eq:mah}
\end{equation}
where $M_0$ is the descendant halo mass at $z=0$, and $\gamma$ and $\delta$ are
two free parameters.  \citet{mcbrideMassAccretionRates2009} found that the
combination, $\gamma-\delta$, to be a useful parameter to characterize the mass
accretion rate at low $z$, since
\begin{equation}
    {{\rm d}\ln M(z)\over {\rm d}\ln(1 + z)} = \gamma - \delta (1 + z) \approx
    \gamma - \delta + \mathcal O(z) \label{eq:acc_rate}
\end{equation}
when $z$ is close to zero. As shown in
Fig.\,\ref{fig:figures/size_evolution_mah_examples}, this functional form can
describe the simulated mass accretion histories reasonably well, but not in
detail. One can use the fitting parameters to describe the overall properties
of a mass accretion history, as we will do here. In principle, the fitting
results can also be used to define a half-mass time as the highest $z$ when the
main progenitor has reached half of the final halo mass. The half-mass time
defined in this way is not identical to $z_{\rm half}$ obtained directly from
the simulation data. We have checked that both definitions give similar
results, and we will use the $z_{\rm half}$ obtained directly from the data in
our presentation. We note that there are other attempts to compress halo merger
trees using state-of-the-art statistical tools
\citep[e.g.][]{forero-romeroCoarseGeometryMerger2009,
obreschkowCharacterizingStructureHalo2020}.

Our study proposes a new method which compresses a halo merger tree into a
linear protohalo size history, and identifies that the most important quantity
is the amplitude, $\log R_c$. We emphasize that the protohalo size history is
correlated with the mass accretion history, but encapsulates additional
information of halo assembly that is missed in the mass accretion history. As
shown in Fig.~\ref{fig:figures/correlation_loga_mah}, the amplitude of the
protohalo size history, $\log R_c$, moderately correlates with the halo mass
assembly time, $z_{\rm half}$, in the sense that smaller protohalos tend to
form earlier. This is expected: at fixed descendant halo mass, smaller
protohalos have deeper gravitational potential well, so that the member halos
can merge with each other in a shorter time scale.
Fig.~\ref{fig:figures/correlation_loga_mah} also shows that $\log R_c$ is in
moderate correlation with $\gamma-\delta$, which approximates the recent mass
accretion rate (see equation~\ref{eq:acc_rate}), in the sense that larger
protohalos tend to experience more rapid accretion at late times. We have also
looked into other features extracted from the mass accretion history and found
that $z_{\rm half}$ exhibits the strongest correlation with $\log R_c$.

The protohalo size history also encapsulates information on halo assembly that
is missed in the mass accretion history. First of all, as one can see from
Fig.~\ref{fig:figures/protohalo_size_correlation_in_mass_bin}, $\log R_c$
strongly correlates with the substructure of descendant halos, while a similar
analysis in Appendix~\ref{sec:results_for_the_mass_accretion_history} shows
that the correlation between the mass accretion history and the substructure of
descendant halos is much weaker. Secondly, $\log R_c$ contains extra
information about the halo concentration, $c$. To demonstrate it, we take the
residual of $\log R_c$ and $c$ conditioned on $z_{\rm half}$ and plot their
relation in Fig.~\ref{fig:figures/correlation_zhalf_ph_size_conc2}. Here one
can see that these two residual quantities are still moderately correlated,
with Spearman's coefficient about 0.3. The last manifestation is in the halo
assembly bias effect, as we will show in \S\,\ref{sec:halo_assembly_bias}.

Finally, it is noteworthy that the mass accretion history and the protohalo
size history captures different stages of halo assembly. The mass accretion
history primarily features the late-time evolution when a dominant progenitor
halo emerges, since the characteristic halo formation times derived from the
mass accretion history, such as the half-mass time and the time when the halo
transits from fast accretion to slow accretion, are below $z\lesssim 2$ for
halos above $10^{13}h^{-1}\rm M_\odot$. On the contrary, the protohalo size
history features the early-time evolution prior to the collapse of the
protohalo, which occurs at $z\sim 2$. Therefore, the combination of these two
quantities might give us a more complete picture of halo assembly.

\section{Halo assembly bias}%
\label{sec:halo_assembly_bias}

\begin{figure*}
    \centering
    \includegraphics[width=1\linewidth]{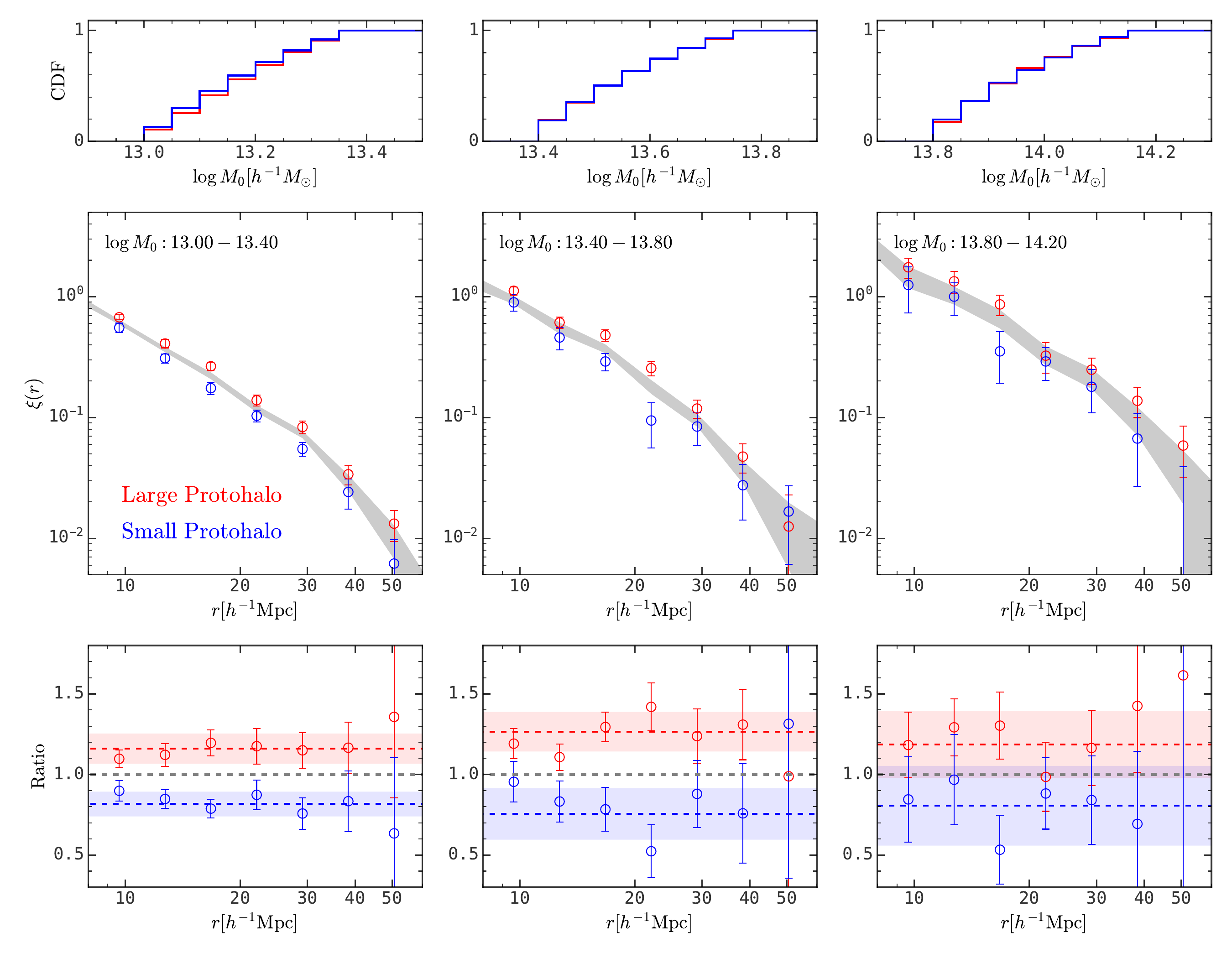}
    \caption{
        Top panels: The auto-correlation functions for dark matter halos
        in three halo mass bins. The gray shaded regions are for all halos in
        each halo mass bin. The red and blue error bars are for subsamples with
        $\Delta A$ above and below zero, respectively. Bottom panels: The
        ratio of the auto-correlation functions between two subsamples and the
        parent sample in each halo mass bin. The shaded regions show the
        average ratio from 10 to 30 $h^{-1}\rm Mpc$ for each subsample. This
        figure shows the difference in the clustering strength for halos with
        large and small protohalos is about $1.2/0.8 - 1\approx 50\%$.
    }%
    \label{fig:figures/assembly_bias}
\end{figure*}

\begin{figure*}
    \centering
    \includegraphics[width=1\linewidth]{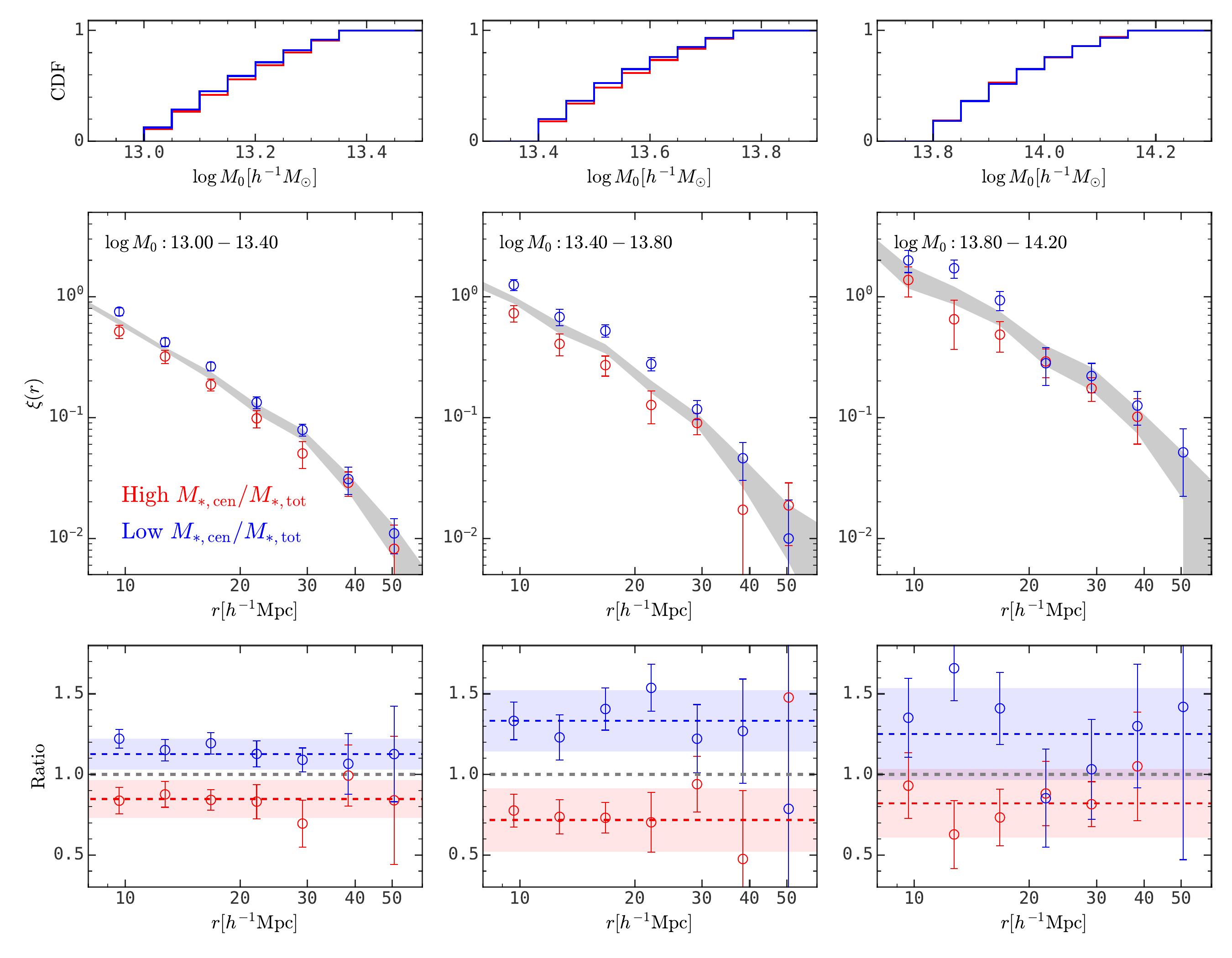}
    \caption{
        Similar to Fig.~\ref{fig:figures/assembly_bias}, except that
        the subsamples are divided according to their $M_{*,\rm cen}/M_{*, \rm
        tot}$.
    }%
    \label{fig:figures/assembly_bias_fcen0}
\end{figure*}

Dark matter halos are biased tracers of the underlying density field, and the
square root of the ratio between the clustering strength of halos and the
underlying density field is defined as the halo bias. The halo bias primarily
depends on halo mass, where massive halos are more clustered
\citep{moAnalyticModelSpatial1996}. In addition, the clustering of dark matter
halos also depends on their secondary properties, including shape, formation
time, concentration, and spin. Such dependence is refered to collectively as the
halo assembly bias \citep[e.g.][]{shethEllipsoidalCollapseImproved2001,
    gaoAgeDependenceHalo2005, wechslerDependenceHaloClustering2006,
    jingDependenceDarkHalo2007, liHaloFormationTimes2008,
maoAssemblyBiasExploring2018}. The existence of halo assembly bias implies the
entanglement of secondary halo properties and their large-scale environments,
and so an analysis of the halo assembly bias can help us
better understand the formation of dark matter halos.

Here we study the assembly bias using the auto-correlation function estimated
with
\begin{equation}
    \xi(r) = \frac{DD(r)}{RR(r)} - 1
\end{equation}
where $DD(r)$ is the pair counts within a distance of $r \pm \delta r$ for the
target sample, and $RR(r)$ is the pair counts within the same distance for the
random sample. We find a strong halo assembly bias effect caused by the
amplitude of the protohalo size history, $\log R_c$. As shown in
Fig.~\ref{fig:figures/assembly_bias}, the auto-correlation function is higher
for halos with larger $R_c$ compared with halos with smaller $R_c$. Here the
subsamples with large and small protohalos are constructed based on their $\log
R_c$ relative to the median relation shown in the right panel of
Fig.~\ref{fig:figures/protohalo_size_depend_m0}. Their descendant halo mass
distributions are nearly identical, as shown in the top panels. Their
auto-correlation functions are presented in the middle panels of
Fig.~\ref{fig:figures/assembly_bias}, where the red and blue error bars are the
results for large and small protohalos, respectively, and the gray shaded
regions are the results for all the halos in the same mass bin. The error bars
in the bottom panels show the ratio between each subsample and the parent
sample. The shaded regions show the mean ratio by averaging the ratios from 10
to 30 $h^{-1}\rm Mpc$, which is an estimate of the square of the relative bias
($b^2=\xi_{\rm subsample}/\xi_{\rm parent}$). Here one can see that the bias
factor between the 50\% of halos with the largest $R_c$ and the 50\% with the
smallest $R_c$ is about $\sqrt{1.2/0.8}- 1\approx 22\%$, which is larger than
the assembly bias caused by the half-mass time. According to the results
obtained in \citet{jingDependenceDarkHalo2007}, the bias factor of the 20\% of
halos with the smallest $z_{\rm half}$ is just about 10\% larger than that the
20\% with the largest $z_{\rm half}$, as is also shown in
Appendix~\ref{sec:results_for_the_mass_accretion_history}. Note that the halo
assembly bias reported here extends to $\gtrsim 40\rm Mpc$, although the error
bars are large due to the limited simulation box size.

Like the mass accretion history, the protocluster size history is not directly
observable. However, as shown in \S\,\ref{sub:relation_to_descendant_halos},
the central-to-total stellar mass ratio, $\log M_{*, \rm cen}/M_{*, \rm tot}$,
is tightly correlated with $R_c$, and so can be served as an observational
proxy. As a proof of concept, Fig.~\ref{fig:figures/assembly_bias_fcen0} shows
the halo assembly bias manifested by $\log M_{*, \rm cen}/M_{*, \rm tot}$. Here
the results are similar to the results we obtained using $R_c$ in
Fig.~\ref{fig:figures/assembly_bias}, indicating that the proxy is reliable.

Actually, halo assembly bias effects based on quantities like $M_{*, \rm
cen}/M_{*, \rm tot}$ have already been detected in observations.
\citet{zuDoesConcentrationDrive2021} used the REDMAPPER cluster catalog
constructed from the SDSS DR8 to study the dependence of the halo assembly bias
caused by the central stellar mass to halo mass ratio, where the halo mass is
controlled using the weak gravitational lensing technique. They found that the
large-scale bias for clusters with lower central stellar mass is 10\% higher
than that of higher central stellar mass from both weak gravitational lensing
and galaxy clustering results \citep{zuStrongConformityAssembly2022}. Under the
assumption that the relation between total stellar mass and halo mass of these
clusters is sufficiently tight \citep{bradshawPhysicalCorrelationsScatter2020},
their detection is similar to the halo assembly bias effect exhibited by the
amplitude of the protohalo size history approximated by the central-to-total
stellar mass ratio. Moreover, they also found a difference in the halo
concentration between subsamples with large and small central stellar masses,
indicating that the halo assembly bias exhibited by the protohalo size and by
the halo concentration are related to each other, as shown in
Fig.~\ref{fig:figures/correlation_loga_conc}.

Finally, we note that the halo assembly bias effect is a manifestation of the
entanglement between halo properties and the large-scale environment. Such
effect is present for the halo concentration, $c=R_{\rm vir}/r_s$, but not for
the half-mass time, $z_{\rm half}$ \citep{jingDependenceDarkHalo2007,
wangEvaluatingOriginsSecondary2021a}, nor the entire mass accretion history
\citep[see][and
Appendix~\ref{sec:neighbor_counts}]{maoAssemblyBiasExploring2018}. This makes
us wonder how the halo concentration ``knows'' the large-scale environment
without the assembly history ``knowing'' it. In this study, we find that the
protohalo size history exhibits a strong halo assembly bias effect. Besides,
Appendix~\ref{sec:neighbor_counts} further demonstrates that paired and
un-paired cluster-size halos have nearly identical mass accretion histories,
but different protohalo size histories, which suggests that the environmental
effects on halo formation must have been lost during the data compression from
halo merger trees to mass accretion histories, and such effects are captured by
the protohalo size history.

\section{Discussion and summary}%
\label{sec:discussion_and_summary}

\subsection{Implications for connecting protoclusters and
clusters}%
\label{sub:implication_on_the_secondary_protocluster_cluster_connection}

\begin{figure*}
    \centering
    \includegraphics[width=1\linewidth]{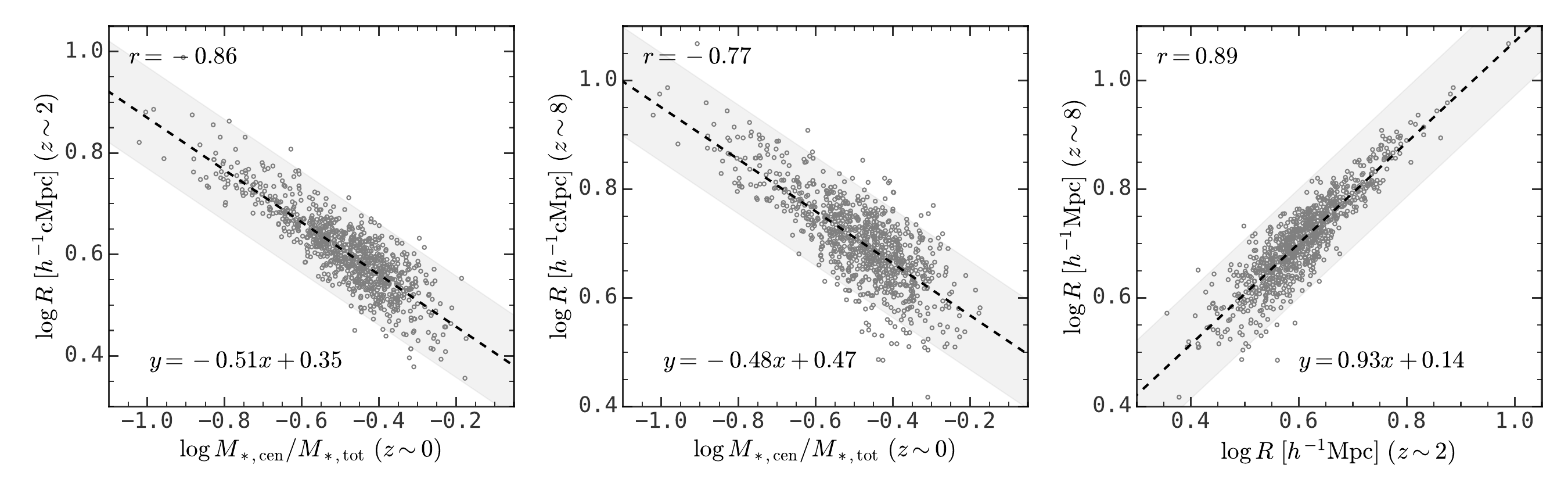}
    \caption{
        Property correlation for protoclusters at $z\sim 2$ and $z\sim 8$ and
        descendant clusters at $z\sim 0$ with descendant halo mass between
        $10^{14}h^{-1}\rm M_\odot$ and $10^{14.2}h^{-1}\rm M_\odot$. Left
        and middle panels: The correlation between the $\log M_{*,\rm
        cen}/M_{\rm *,tot}$ of descendant clusters and the size of their
        protohalo at $z\sim 2$ and $z\sim 8$, respectively. Right panel:
        The correlation between the sizes of protohalos at $z\sim 2$ and $z\sim
        8$. The black dashed lines are the linear fitting results and the
        shaded regions are the $\pm 0.1$dex regions. Pearson's correlation
        coefficients are presented at the top-left corner of each panel. This
        figure demonstrates that the protohalo size serves as an excellent
        secondary property to connect protoclusters across cosmic time and
        their descendant clusters through their $\log M_{*, \rm cen}/M_{*, \rm
        tot}$.
    }%
    \label{fig:figures/protocluster_secondary}
\end{figure*}

 A protocluster, by definition, is a set of dark matter halos, as well as their
 associated baryons, at $z> 0$ that will eventually collapse into a common
 cluster-size dark matter halo at $z=0$. In other words, protoclusters are
 protohalos whose descendant halo mass is $\gtrsim 10^{14}h^{-1}\rm M_\odot$.
 Such objects are well-defined in simulations where we can track their
 evolution. However, it is non-trivial to identify protoclusters in observation
 due to the difficulty in predicting the fate of those high-$z$ objects. A
 feasible way is to characterize true protoclusters in simulations with a few
 features, such as the galaxy number overdensity, and find objects in the
 observation that exhibit similar properties
 \citep[e.g.][]{chiangANCIENTLIGHTYOUNG2013}. This kind of method is
 straightforward but their performance is difficult to assess since they rely
 on the visual inpsection. A more advanced way is to develop an automatic
 structure finding algorithm, which can be trained in simulations and applied
 to mock surveys for performance assessments
 \citep[e.g.][]{starkProtoclusterDiscoveryTomographic2015,
 wangFindingProtoclustersTrace2021}.

How to build a reliable connection between protoclusters and present-day
clusters is, therefore, an essential step in understanding the formation of
clusters, and the evolution of cluster galaxies. Conventionally, such
connections are built using masses of protoclusters and descendant clusters.
For each identified protocluster, the mass can be estimated according to
proxies such as the number overdensity of tracers and the total mass within the
protocluster. However, the mass alone cannot capture the diversity of
protocluster, as shown in Fig.~\ref{fig:figures/examples_of_pc_diff_z2}
\citep[see also][]{lovellCharacterisingIdentifyingGalaxy2018}.

We find that the protocluster size, which is defined in the same manner as the
protohalo size, is an excellent secondary property to connect protoclusters
across cosmic time and their descendant clusters at $z\sim 0$. The left and
middle panels of Fig.~\ref{fig:figures/protocluster_secondary} show the
correlation between the central-to-total stellar mass ratio of descendant
clusters and the protocluster size at $z\sim 2$ and $z\sim 8$, respectively.
Here one can see that these correlations are tight, with Pearson's correlation
coefficients as high as $\sim 0.86 (0.77)$ for $z\sim 2 (8)$. The right panel
shows the correlation between the sizes of protoclusters at $z\sim 2$ and
$z\sim 8$, and the correlation coefficient is about $0.9$.

Thus, the protocluster size can be used to refine the connections of
protoclusters across cosmic time and to local galaxy clusters. This refinement
may be referred to as the secondary cluster-protocluster connection.
Observationally, the usefulness of this refinement relies on the fact that the
protocluster size and the central-to-total stellar mass ratio of descendant
halos are tightly correlated and that galaxy groups and protoclusters can be
reliably identified from observations \citep{yangGalaxyGroupsSDSS2007,
    wangIdentifyingGalaxyGroups2020, yangExtendedHalobasedGroup2021,
liGroupsProtoclusterCandidates2022}. However, our findings also pose a
challenge to the identification of protoclusters from observation. As we have
shown, the size difference between protoclusters of similar descendant mass can
be as high as $\sim 0.5$ dex. This corresponds to a factor of $10^{0.5\times
3}\approx 30$ in volume, and suggests that protocluster finders based on
overdensities within a fixed aperture may produce a biased protocluster
catalog. Clearly, for a density-based protocluster finder, such bias needs to
be quantified using realistic mock catalogs
\citep{starkProtoclusterDiscoveryTomographic2015,
wangFindingProtoclustersTrace2021}.

\subsection{Summary}%
\label{sub:summary}

We propose a novel method to characterize the assembly of
massive dark matter halos using the protohalo size history.
Our main findings are summarized as follows:
\begin{enumerate}

    \item The protohalo size history exhibits two-stage evolution: an
        early-time static phase and a late-time collapsing phase. These
        features can be captured by a double power-law function with a
        characteristic redshift at $z=2$ in the halo mass range between
        $10^{13}h^{-1}\rm M_\odot$ and $10^{15}h^{-1}\rm M_\odot$. The
        late-time slope $\alpha$ and the early-time slope $\beta$ are nearly
        independent of the descendant halo mass, while the amplitude $\log R_c$
        positively correlates with the descendant halo mass with a slope of $\sim
        0.39$ (see Figs.~\ref{fig:figures/size_evolution_mah_examples} and
        \ref{fig:figures/protohalo_size_depend_m0}).

    \item At a given descendant halo mass, the amplitude of the protohalo size
        history, $\log R_c$, strongly correlates to the central-to-total stellar
        mass ratio, $\log M_{*, \rm cen}/M_{*, \rm tot}$, of the descendant
        halos, with smaller protohalos evolving to descendant halos
        that are more dominated by central galaxies. There is also a moderate
        correlation between the late-time slope and $\log M_{*, \rm cen}/M_{*,
        \rm tot}$, in the sense that more rapid collapsing produces a more dominating
        central galaxy (see Fig.~\ref{fig:figures/protohalo_size_evolution}
        and Appendix~\ref{sec:results_for_the_mass_accretion_history}).

    \item The protohalo size history correlates with the mass accretion
        history, but also encapsulates critical information about halo assembly that is
        missed by the mass accretion history.
        This is reflected in its tight
        correlation with the central-to-total stellar mass ratio, $\log M_{*,
        \rm cen}/M_{*, \rm tot}$, and its correlation with halo
        concentration when $z_{\rm half}$ is fixed (see
        Figs.~\ref{fig:figures/correlation_loga_mah},
        \ref{fig:figures/protohalo_size_correlation_in_mass_bin}, and
        \ref{fig:figures/correlation_zhalf_ph_size_conc2}).

    \item The amplitude of the protohalo size exhibits a strong halo assembly
        bias effect, in that descendant halos of a given mass
        with larger protohalo sizes are more strongly correlated.
        A similar assembly bias effect is
        found for the central-to-total stellar mass ratio of descendant halos,
        due to its strong correlation to the protohalo size. However, the halo
        assembly bias exhibited by the half-mass time is much weaker at the massive
        end. This indicates that the information about halo assembly bias is
        lost during the data compression from halo merger trees to the mass
        accretion histories, and this information is captured by the protohalo
        size histories (see Figs.~\ref{fig:figures/assembly_bias} and
        \ref{fig:figures/assembly_bias_fcen0} and
        Appendix~\ref{sec:results_for_the_mass_accretion_history}).

    \item The sizes of protoclusters at $z\sim 2$ to $z\sim 8$ are all strongly
        correlated with the central-to-total stellar mass ratio of their
        descendant clusters at $z\sim 0$. The sizes of protoclusters
        at different redshifts are also strongly correlated to each other, which indicates the protocluster size is
        a useful quantity to link protoclusters at high-$z$
        to their descendants at $z\sim 0$
        (see Fig.~\ref{fig:figures/protocluster_secondary}).

\end{enumerate}

Our results suggest that the protohalo size history may provide
a new avenue to study the halo assembly history and its relation to galaxy
formation and evolution.  The amplitude of the protohalo size history has a
reliable observational proxy, which is the central-to-total stellar mass ratio.
In contrast, observational proxies for the commonly-used mass accretion
histories are not as reliable, especially for massive halos
\citep{wangLateformedHaloesPrefer2023}. The protohalo size history also
encapsulates information about the halo assembly that is missed in the mass
accretion history, as is manifested by the halo assembly bias effect. Finally,
the protohalo size can be used as a secondary parameter, in addition to mass,
to link protoclusters across cosmic time, and to link protoclusters to
descendant clusters at $z\sim 0$. This will help us to better understand the
assembly of galaxy clusters and the evolution of their member galaxies, when
applied to the existing and upcoming high-$z$ galaxy surveys, such as MAMMOTH
\citep{caiMappingMostMassive2016, caiMappingMostMassive2017}, COSMOS-Webb
\citep{caseyCOSMOSWebOverviewJWST2022}, PFS
\citep{greenePrimeFocusSpectrograph2022a}, and MOONS
\citep{maiolinoMOONRISEMainMOONS2020}.

\section*{Acknowledgements}

The authors acknowledge the Tsinghua Astrophysics High-Performance Computing
platform at Tsinghua University for providing computational and data storage
resources that have contributed to the research results reported within this
paper. KW and YP are supported by the National Science Foundation of China
(NSFC) Grant No. 12125301, 12192220, 12192222, and the science research grants
from the China Manned Space Project with NO. CMS-CSST-2021-A07. HW is supported
by NSFC Grant No. 12192224.

The computation in this work is supported by the HPC toolkit \specialname[HIPP]
\citep{hipp}, IPYTHON \citep{perezIPythonSystemInteractive2007}, MATPLOTLIB
\citep{hunterMatplotlib2DGraphics2007}, NUMPY
\citep{vanderwaltNumPyArrayStructure2011}, SCIPY
\citep{virtanenSciPyFundamentalAlgorithms2020}, ASTROPY
\citep{astropy:2013, astropy:2018, astropy:2022}. This research made use of
NASA’s Astrophysics Data System for bibliographic information.

\section*{Data availability}

The data underlying this article will be shared on reasonable request to the
corresponding author.

\bibliographystyle{mnras}
\bibliography{bibtex.bib}

\appendix

\section{Protohalo sizes using different halo mass limits}%
\label{sec:protohalo_sizes_with_different_halo_mass_limits}

\begin{figure*}
    \centering
    \includegraphics[width=1\linewidth]{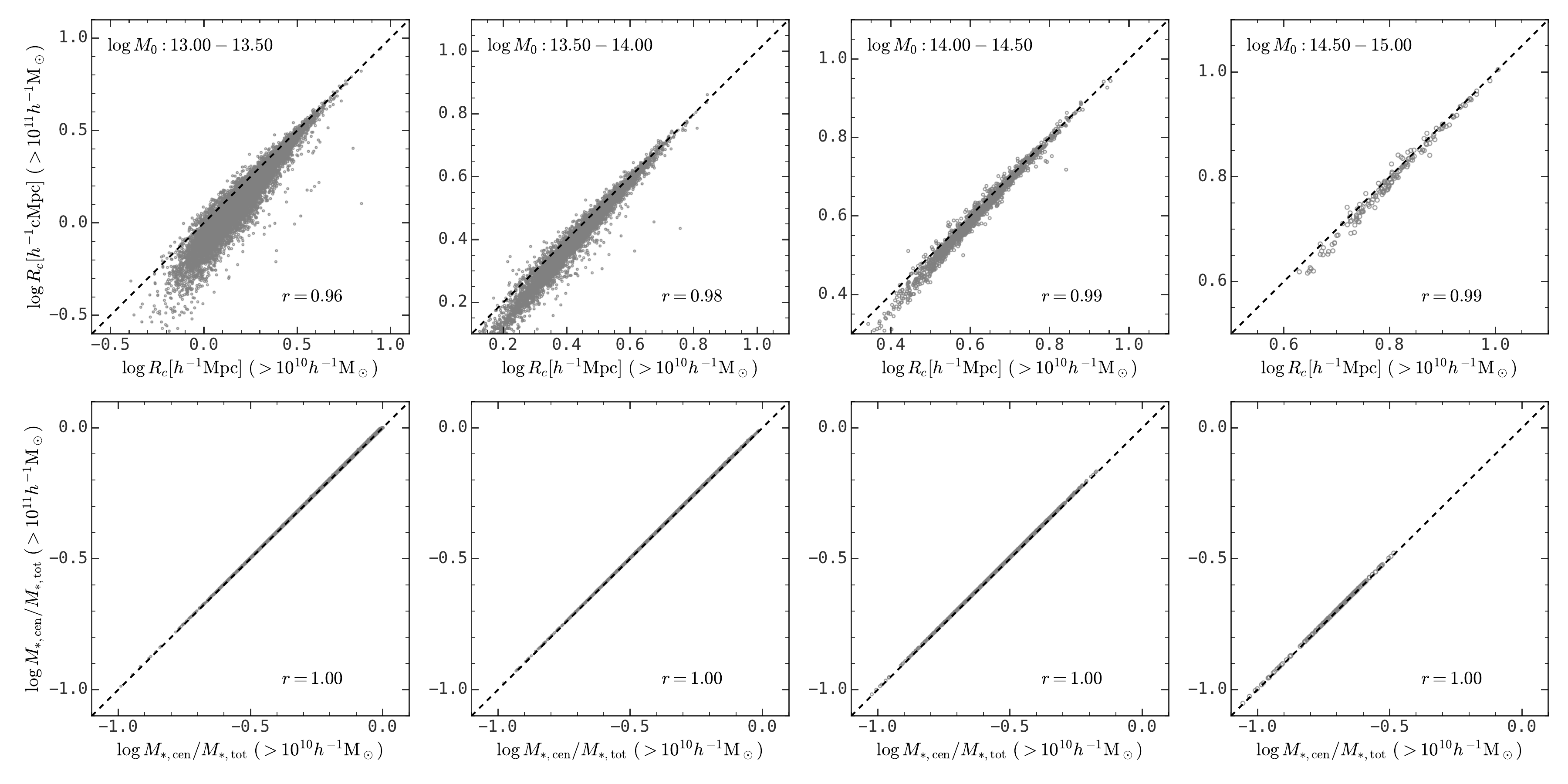}
    \caption{
        The comparison of the amplitude of the protohalo size history (top
        panels) and the central-to-total stellar mass ratio (bottom
        panels) with the halo mass limit as $10^{10}h^{-1}\rm M_\odot$
        ($x$-axis) and $10^{11}h^{-1}\rm M_\odot$ ($y$-axis). Spearman's rank
        correlation coefficients are presented on each panel. This figure shows
        that a higher halo mass limit causes an under-estimation of the
        amplitude of the protohalo size history, $\log A$, for halos with low
        mass and small protohalo size, but little damage to the rank of $\log
        R_c$. The impact on the central-to-total stellar mass ratio, $\log
        M_{*, \rm cen}/M_{*, \rm tot}$ is negligible.
    }%
    \label{fig:figures/compare_psh_hm_limit}
\end{figure*}

Here we examine the impact of the halo mass limit used to estimate the protohalo
size and the central-to-total stellar mass ratio of descendant halos.

The top panels of Fig.~\ref{fig:figures/compare_psh_hm_limit} show the scatter
plot of the amplitude of the protohalo size history, i.e. $\log R_c$, estimated
using progenitor halos above $10^{10}h^{-1}\rm M_\odot$ (the limit adopted in
the main  body of the paper) and $10^{11}h^{-1}\rm M_\odot$, and Spearman's
rank correlation coefficients are presented in each panel. Here one can see
that a higher halo mass limit will cause an underestimation of the protohalo
size. This is expected since massive halos are more biased and prefer to live
in central parts of protohalos.  This underestimation is moderate for
group-size descendant halos and negligible for cluster-size descendant halos.
Most importantly, the rank correlation coefficient between different halo mass
limits is $\sim 0.96$ in the lowest descendant halo mass bin and even higher
for more massive bins,  indicating that such underestimation effects do not
alter the relative ranks of $\log R_c$ for these halos.

The bottom panels of Fig.~\ref{fig:figures/compare_psh_hm_limit} shows the
central-to-total stellar mass ratio of descendant halos, i.e. $\log M_{*, \rm
cen}/M_{*, \rm tot}$, using all subhalos with peak halo mass above
$10^{10}h^{-1}\rm M_\odot$ and $10^{11}h^{-1}\rm M_\odot$, respectively.
There is nearly no difference between these two quantities, indicating again that
subhalos with low peak halo mass have a negligible contribution to the total stellar mass.

\section{Protohalo sizes using different centering methods}%
\label{sec:centering}

\begin{figure*}
    \centering
    \includegraphics[width=1\linewidth]{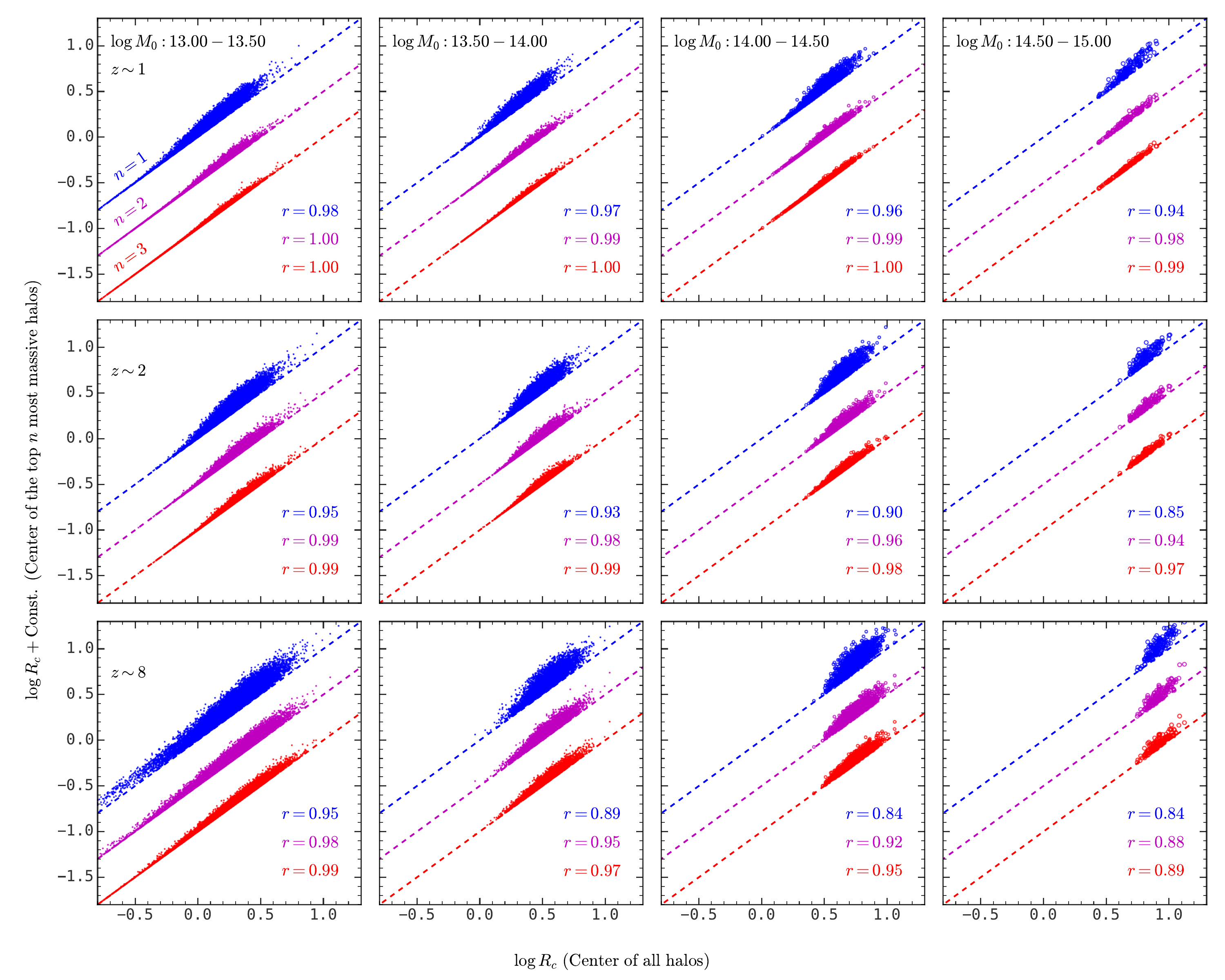}
    \caption{
        The comparison of protohalo sizes calculated with different centers.
        The $x$-axis is the protohalo size calculated with the center of mass.
        The $y$-axis is the protohalo size calculated with the center of mass
        for $n$ most massive halos in the protohalo, where $n$=1 (blue), 2
        (magenta), and 3 (red). Spearman's correlation coefficients between
        two protohalo sizes are shown on each panel in the corresponding color.
        This figure demonstrates that centering with a few most massive halos
        will overestimate the protohalo size, and the overestimation is larger
        for situations without dominating halos, such as large protohalos,
        high-$z$ protohalos, and protohalos with massive descendant halo mass.
        Nevertheless, the rank of the protohalo size is largely preserved,
        which is manifested by the high values of Spearman's correlation coefficients.
    }%
    \label{fig:figures/compare_centering}
\end{figure*}

Here we examine the impact of the protohalo centering on the protohalo size
calculations. To this end, we use four different methods to determine the protohalo center
$\mathbf x_{\rm cen}$:
\begin{enumerate}
    \item using the mass of center of all halos above $10^{10}h^{-1}\rm
        M_\odot$ in each protohalo, which is used in the main body of the paper;
    \item using the position of the most massive halo above $10^{10}h^{-1}\rm
        M_\odot$ in each protohalo;
    \item using the mass of center of top-2 most massive halos above
        $10^{10}h^{-1}\rm M_\odot$ in each protohalo;
    \item using the mass of center of top-3 most massive halos above
        $10^{10}h^{-1}\rm M_\odot$ in each protohalo.
\end{enumerate}
In the latter two cases, if the number of halos above $10^{10}h^{-1}\rm M_\odot$
in the protohalo is smaller than the required number, we use all the
available halos.

Fig.~\ref{fig:figures/compare_centering} shows the scatter of the protohalo
size estimated with different centering methods in four halo mass bins and
three redshift snapshots. In each panel, the $x$-axis is the protohalo size
calculated with the first centering method, and the $y$-axis shows those with
the other three methods with vertical shifts for clarity. The dashed lines
in each panel are the one-to-one reference lines. Spearman's rank correlation
coefficients for the protohalo sizes from different centering methods are
presented on each panel. Here one can see that the first centering method
produces the minimal protohalo size by design, as the other three methods will
overestimate the sizes. Moreover, this overestimation effect has some general
trends. First of all, the amplitude of the overestimation decreases with $n$,
which is expected. Secondly, the protohalo sizes are more overestimated for
larger protohalos, which can be seen from that the deviation from the reference
line is larger for larger $x$-values. Thirdly, such overestimation effects are
larger for high-$z$ protohalos, which can be seen from the decreasing rank
correlation coefficients with increasing redshift. Similarly, the
overestimation effect is larger for protohalos with massive descendant halos.
In summary, the overestimation effect is stronger in situations where the
protohalos are less likely to be dominated by a few massive halos.
Nevertheless, the rank of of the protohalo size is largely preserved, since the
rank correlation coefficients are $\gtrsim 0.85$ even when only the most
massive halo is used.

\section{Protohalo sizes using different subhalo membership definition}%
\label{sec:protohalo_sizes_using_different_subhalo_definition}

\begin{figure*}
    \centering
    \includegraphics[width=1\linewidth]{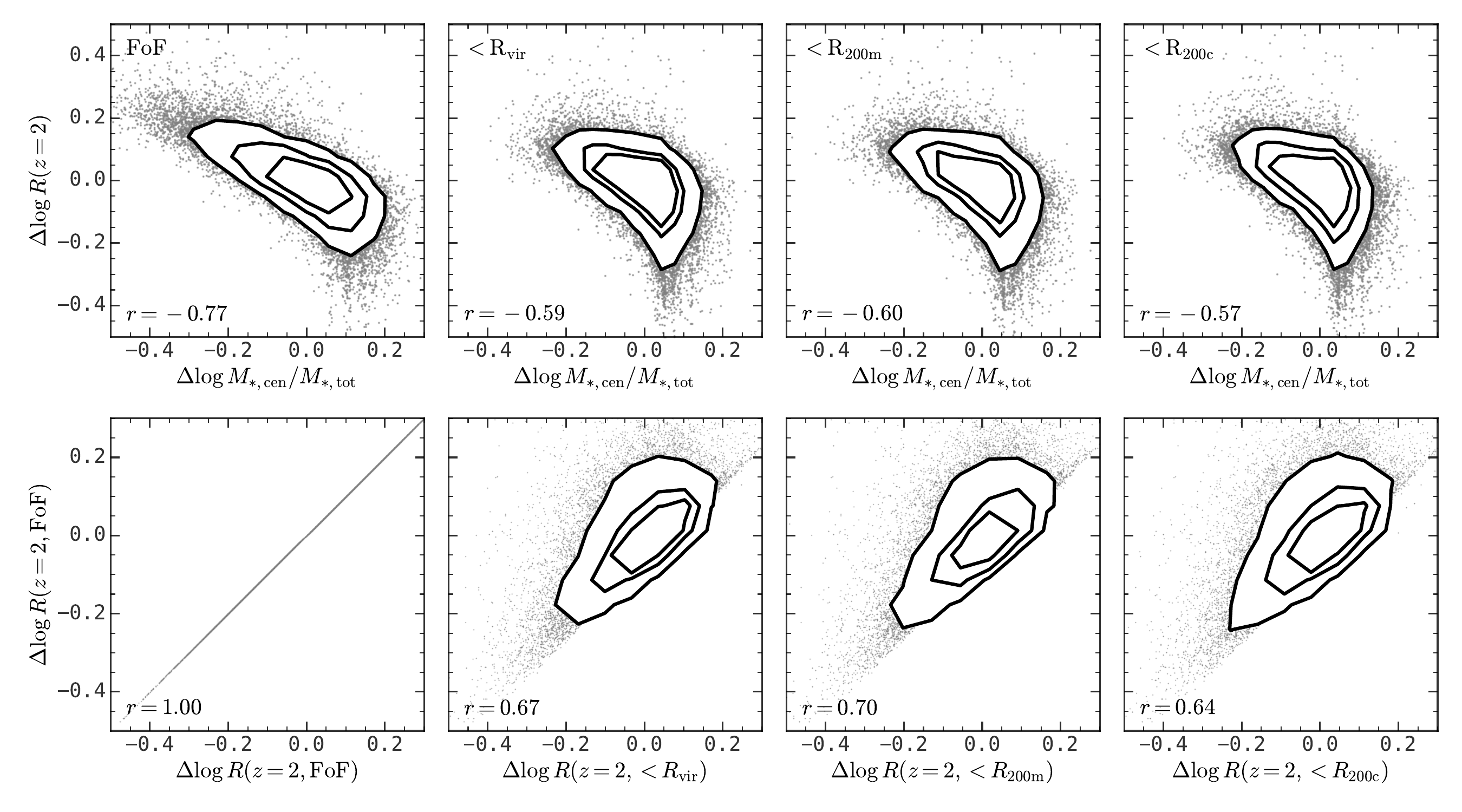}
    \caption{
        The top panels show the relation between the residual protohalo size at
        $z=2$ and the residual central-to-total stellar mass ratio at $z=0$ for
        different subhalo membership definitions for all descendant halos above
        $10^{13}h^{-1}\rm M_\odot$. Similarly, the bottom panels show the
        comparison of the protohalo sizes at $z=2$. The contour lines show
        encloses 50\%, 70\%, and 90\% of all halos. The first column includes
        all subhalos in each FoF halo, and the second column includes only
        subhalos within $R_{\rm vir}$. Similarly, the third and fourth columns
        include only subhalos within $R_{\rm 200m}$ and $R_{\rm 200c}$,
        respectively.
    }%
    \label{fig:figures/different_radius}
\end{figure*}

In the main text, we define a protohalo as the collection of progenitor halos
of all subhalos in the descendant FoF halo. Alternatively, we can define a
protohalo as all of the progenitor halos that will end up in descendant
subhalos that are enclosed within the virial radius of the descendant halo at
$z=0$. Here we considered three different common halo radius defined with
different density threshold values: $R_{\rm vir}$ defined with the threshold
density in \citet{bryanStatisticalPropertiesXRay1998}, $R_{\rm 200m}$ defined
with the threshold density as $200\Omega_{\rm m}\rho_{\rm crit}$, and $R_{\rm
200c}$ defined with the threshold density as $200\rho_{\rm crit}$. The bottom
panels of Fig.~\ref{fig:figures/different_radius} show the comparison of
protohalo sizes at $z=2$ using different definitions. Note that the dependence
of protohalo size on descendant halo mass is eliminated by taking the residual
with respect to the median $\log R-\log M_0$ relation. The protohalo sizes
defined with additional descendants' halo-centric distance constraint are
smaller than that defined with all descendant subhalos in the descendant FoF
halo. This is expected since the progenitor halos of the subhalos on the
descendant halo's outskirt prefer to locate in the outer region of the
protohalo, so that excluding these progenitor halos will consequently reduce
the protohalo size. Nevertheless, these protohalo sizes with different
definitions still strongly correlate to each other.

The top panels of Fig.~\ref{fig:figures/different_radius} show the relation
between the protohalo size and the central-to-total stellar mass of descendant
halos with the dependence on halo mass eliminated by taking the deviation from
the median value in narrow halo mass bins. Note that the calculation of
$M_{*,\rm tot}$ only includes subhalos within the corresponding radius. After
eliminating subhalos out of halo radius, the correlation coefficient between
the protohalo size and the central-to-total stellar mass ratio is reduced but
still as high as $\approx 0.6$.

\section{Results for the IllustrisTNG simulation}%
\label{sec:results_for_the_illustrstng_simulation}

\begin{figure*}
    \centering
    \includegraphics[width=0.9\linewidth]{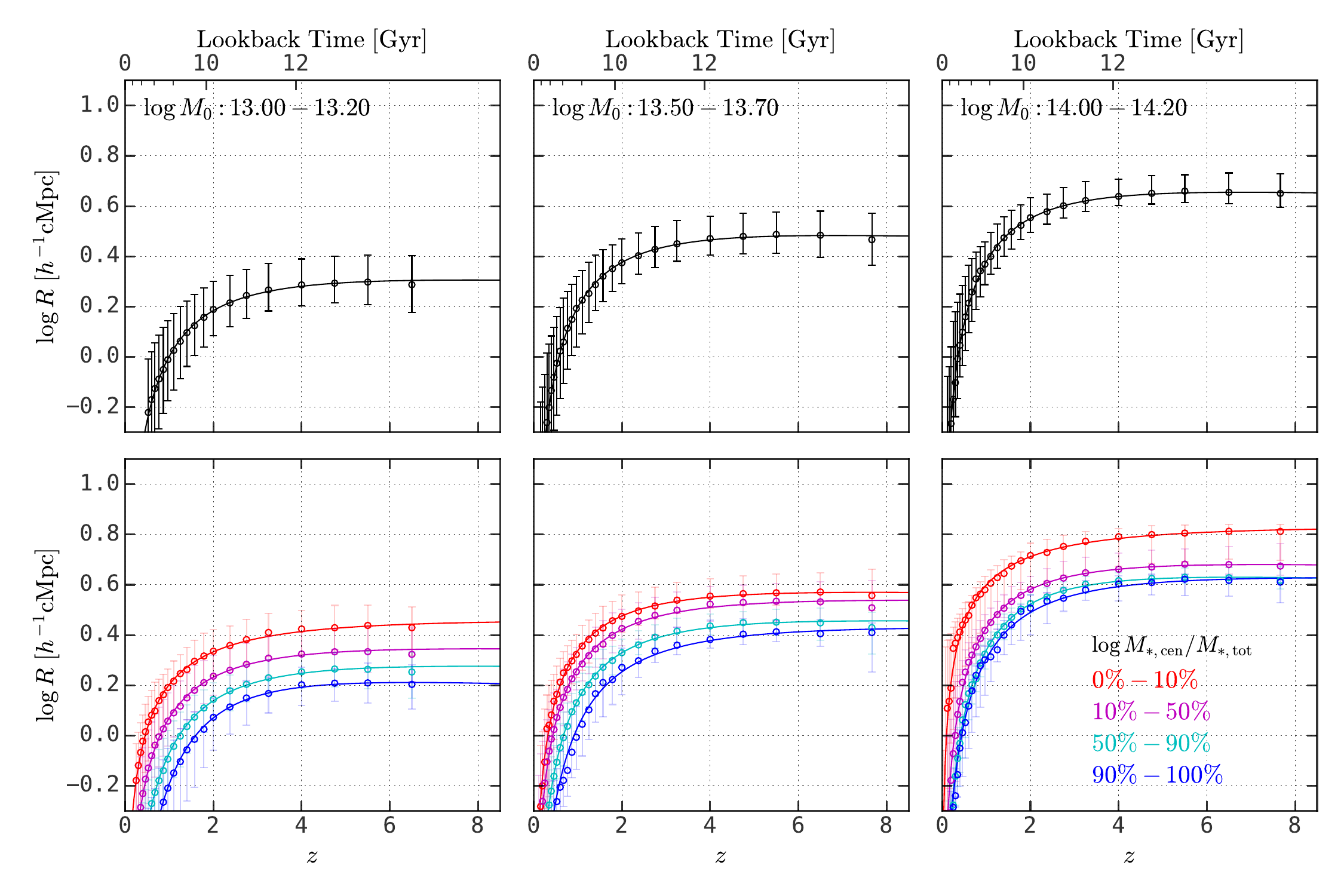}
    \caption{
        Similar to Fig.~\ref{fig:figures/protohalo_size_evolution}, except for
        the TNG300 simulation.
    }%
    \label{fig:figures/protohalo_size_evolution_tng}
\end{figure*}

\begin{figure*}
    \centering
    \includegraphics[width=0.9\linewidth]{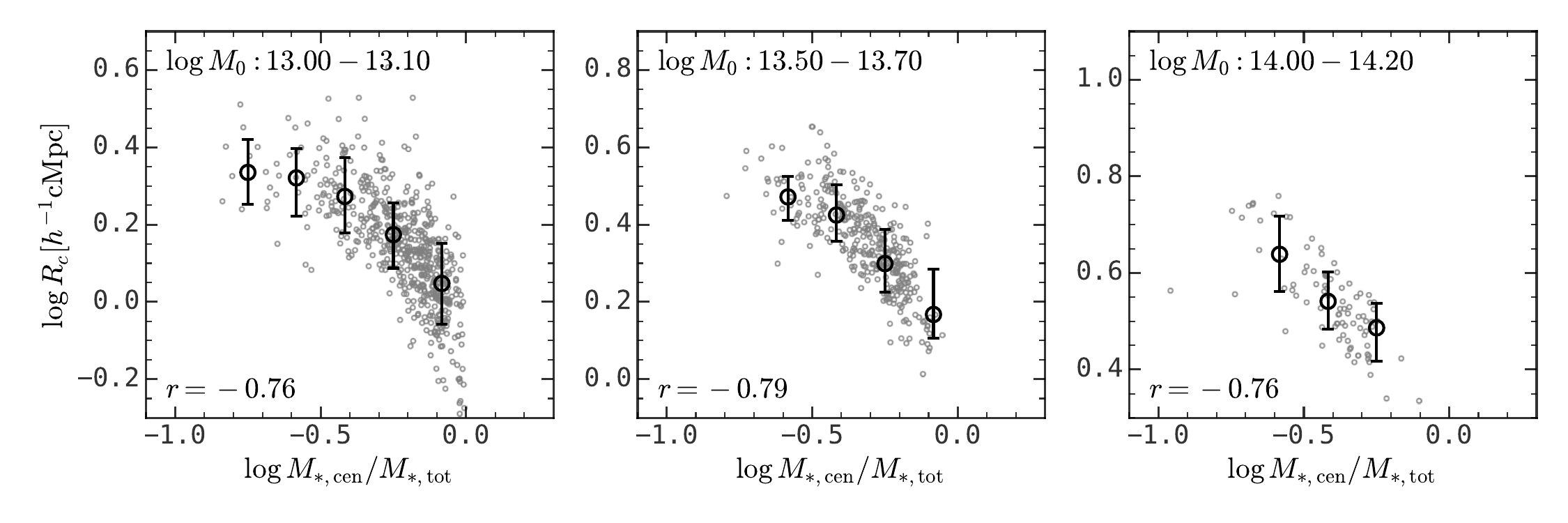}
    \caption{
        Similar to the bottom panels of
        Fig.~\ref{fig:figures/protohalo_size_correlation_in_mass_bin}, except
        for the TNG300 simulation.
    }%
    \label{fig:figures/protohalo_size_correlation_in_mass_bin_tng}
\end{figure*}

\begin{figure*}
    \centering
    \includegraphics[width=0.8\linewidth]{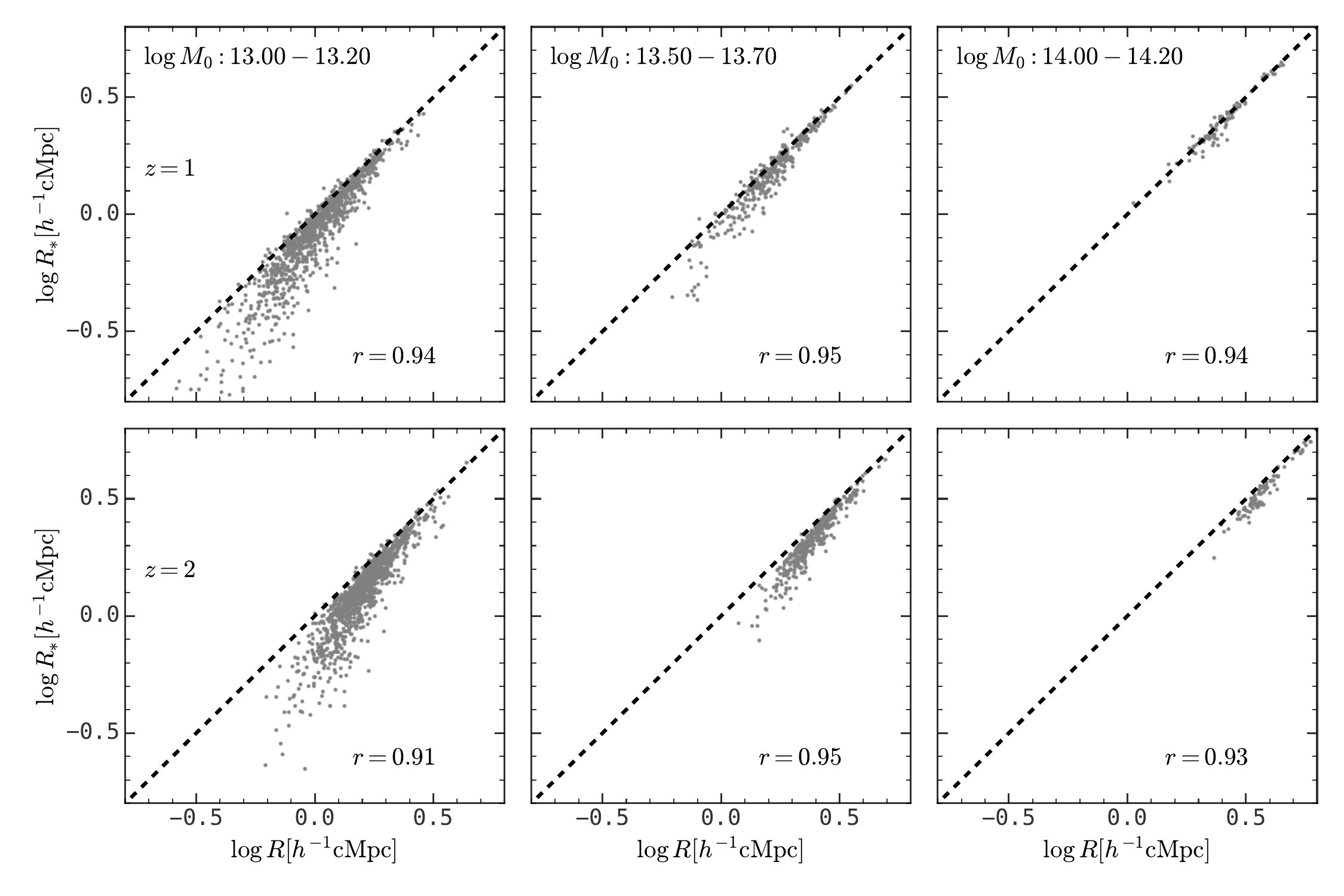}
    \caption{
        The comparison of protohalo size calculated with halo mass ($x$-axis)
        and stellar mass ($y$-axis) in TNG300 at $z=1$ and $z=2$. Spearman's
        rank correlation coefficients are labelled at the lower right corner of
        each panel. This figure demonstrates that the protohalo sizes
        calculated using halo mass and stellar mass are highly correlated to
        each other, despite that the size calculated using stellar mass is
        systematically smaller than that of halo mass for small protohalos.
    }%
    \label{fig:figures/compare_psh_sm_tracer_tng300}
\end{figure*}

The IllustrisTNG project comprises various cosmological galaxy formation
simulations with different box sizes and resolutions \citep{Marinacci_2018,
    Naiman_2018, Nelson_2018, Nelson_2019, Springel_2018, Pillepich_2018a,
Pillepich_2018b}. Here we use the one with the largest box size, which is
called the TNG300 simulation, for better statistics. TNG300 simulates the
formation and evolution of dark matter halos and galaxies in a box with a side
length of $205h^{-1}\rm Mpc$. The dark matter particle mass is about $5.9\times
10^7\rm M_\odot$, and the mass for each gas particle is about $1.1\times
10^7\rm M_\odot$.

Fig.~\ref{fig:figures/protohalo_size_evolution_tng} shows the protohalo size
evolution as a function of redshift in three descendant halo mass bins. The
evolution is very similar to that obtained from the ELUCID simulation (see
Fig.~\ref{fig:figures/protohalo_size_evolution}. The bottom panels show the
protohalos size evolution binned with the central-to-total stellar mass ratio
of descendant halos, $\log M_{*, \rm cen}/M_{*, \rm tot}$. Here one can see
that protohalos with the top-10\% and bottom-10\% $M_{*, \rm cen}/M_{*, \rm
tot}$ values differ in their size by 0.2-0.3 dex, despite the fact that the
$1-\sigma$ range of the protohalo size is only 0.1-0.2 dex (see top panels). In
addition. Fig.~\ref{fig:figures/protohalo_size_correlation_in_mass_bin_tng}
shows the correlation between $\log R_c$ and $\log M_{*, \rm cen}/M_{*, \rm
tot}$ in three halo mass bins, as well as Spearman's rank correlation
coefficient, which is about 0.8. These results show that the tight correlation
between the protohalo size and the central-to-total stellar mass ratio of
descendant halos is also present in the IllustrisTNG simulation, where the
stellar mass in each halo is obtained through hydrodynamical simulation instead
of empirical modeling.

Fig.~\ref{fig:figures/compare_psh_sm_tracer_tng300} shows the comparison of
protohalo size calculated with equation~(\ref{eq:ph_size}) and that weighted by
the stellar mass of all progenitor galaxies, which is
\begin{equation}
    R_* = {\sqrt{\sum m_{*, i}\|\mathbf x_{*, i} - \mathbf x_{\rm cen,
    *}\|^2\over \sum m_{*, i}}}, ~~~\mathbf x_{\rm cen, *} \equiv {\sum_im_{*,
i}\mathbf x_{*, i}\over \sum_i m_{*,i}} \label{eq:ph_size_sm}
\end{equation}
where $m_{*, i}$ and $\mathbf x_{*, i}$ are the stellar mass and position of
the $i$-th progenitor galaxy. Here one can see that the protohalo sizes
calculated using these two different ways are highly correlated to each other,
despite that the size calculated using stellar mass is systematically smaller
than that of halo mass for small protohalos, which is caused by the
non-constant stellar mass to halo mass ratio
\citep{behrooziUNIVERSEMACHINECorrelationGalaxy2019}.

\section{Results for the mass accretion history}%
\label{sec:results_for_the_mass_accretion_history}

\begin{figure*}
    \centering
    \includegraphics[width=0.8\linewidth]{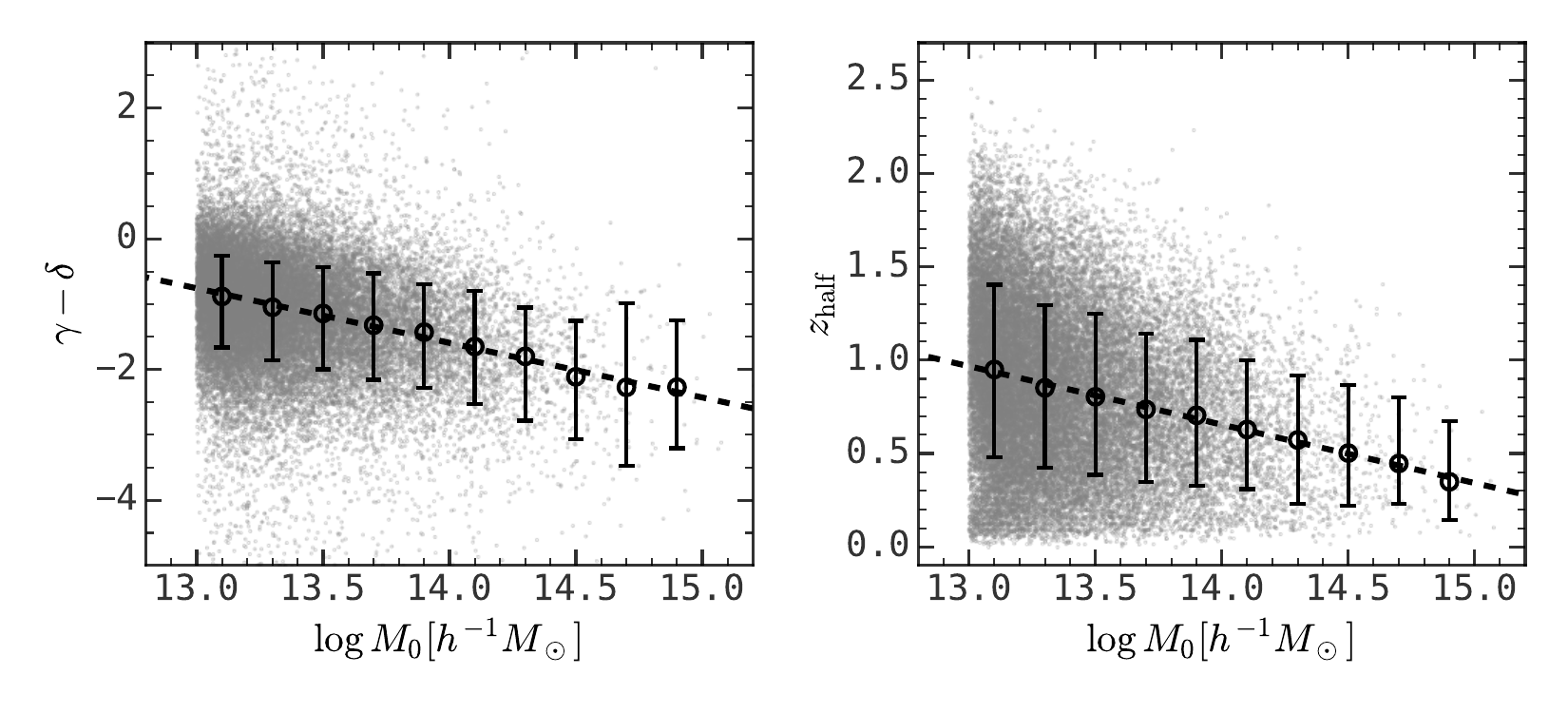}
    \caption{
        Similar to Fig.~\ref{fig:figures/protohalo_size_depend_m0}, except for
        $\gamma-\delta$ and $z_{\rm half}$.
    }%
    \label{fig:figures/mah_depend_m0}
\end{figure*}

\begin{figure*}
    \centering
    \includegraphics[width=1\linewidth]{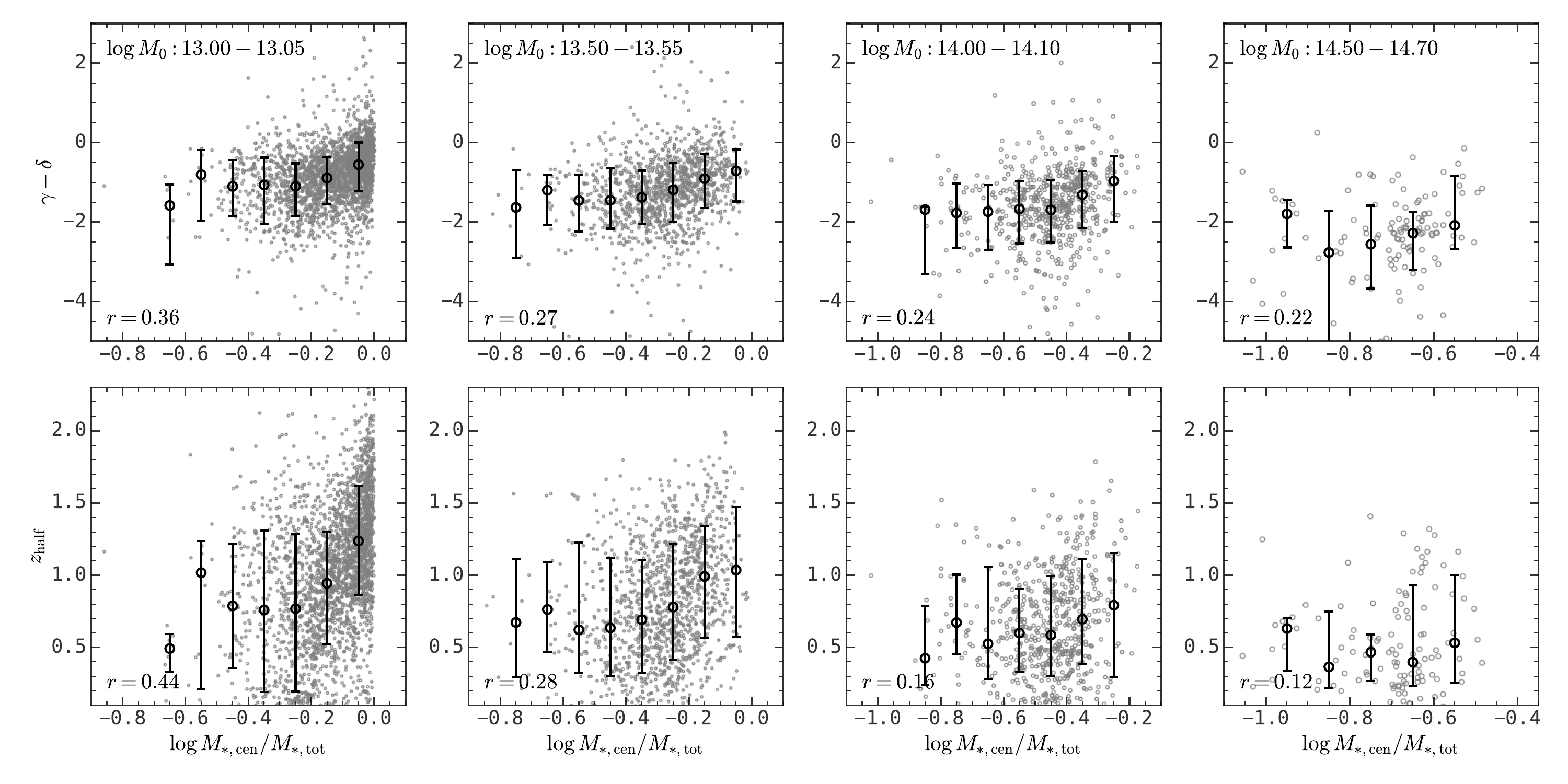}
    \caption{
        Similar to
        Fig.~\ref{fig:figures/protohalo_size_correlation_in_mass_bin}, except
        for $\gamma-\delta$ and $z_{\rm half}$.
    }%
    \label{fig:figures/mah_correlation_in_mass_bin}
\end{figure*}

\begin{figure*}
    \centering
    \includegraphics[width=0.9\linewidth]{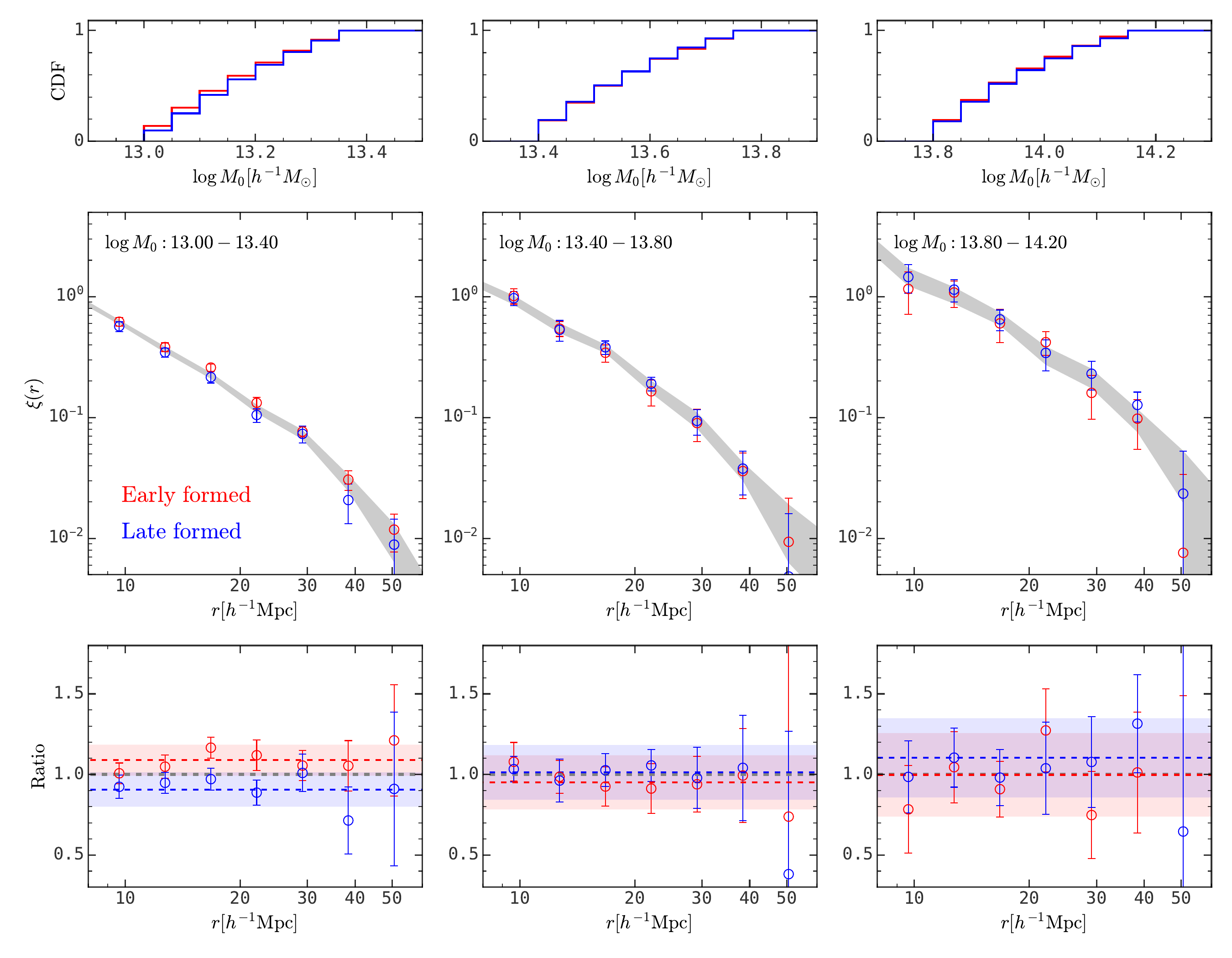}
    \caption{
        Similar to Fig.~\ref{fig:figures/assembly_bias}, except that subsamples
        here are divided according to their $z_{\rm half}$. This figure shows
        that the halo assembly bias from the halo formation time is negligible
        for massive halos.
    }%
    \label{fig:figures/assembly_bias_zhalf}
\end{figure*}

Here we perform similar studies on the mass accretion history as we did on the
protohalo size history in
\S\,\ref{sec:protohalo_size_evolution_and_its_relation_to_descendant_halos} and
\S\,\ref{sec:halo_assembly_bias}.

Fig.~\ref{fig:figures/mah_depend_m0} shows the dependence on descendant halo
mass for $\gamma-\delta$ in equation~\ref{eq:mah} and the half-mass time
$z_{\rm half}$, and the relations to the central-to-total stellar mass ratio,
i.e. $\log M_{*, \rm cen}/M_{*, \rm tot}$, are presented in
Fig.~\ref{fig:figures/mah_correlation_in_mass_bin}. Here one can see that both
$\gamma - \delta$ and $z_{\rm half}$ moderately correlate with $\log M_{*, \rm
cen}/M_{*, \rm tot}$ for group-size halos. Halos with more dominating central
galaxies tend to form earlier and have lower recent accretion rates.  The
correlation becomes negligible for cluster-size halos, which is consistent with
the results in \citet{wangLateformedHaloesPrefer2023}. Neither of the
parameters describing the mass accretion history has as a tight relation to
$\log M_{*, \rm cen}/M_{*, \rm tot}$ as the protohalo size history shown in
Fig.~\ref{fig:figures/protohalo_size_correlation_in_mass_bin}.

Fig.~\ref{fig:figures/assembly_bias_zhalf} shows the halo assembly bias effect
exhibited by the half-mass time $z_{\rm half}$. The signal is rather
weak, which is consistent with the results obtained in previous studies
\citep[see][]{gaoAgeDependenceHalo2005, jingDependenceDarkHalo2007,
maoAssemblyBiasExploring2018}.

\section{The dependence of assembly histories on descendants' neighbor counts}%
\label{sec:neighbor_counts}

\begin{figure*}
    \centering
    \includegraphics[width=0.9\linewidth]{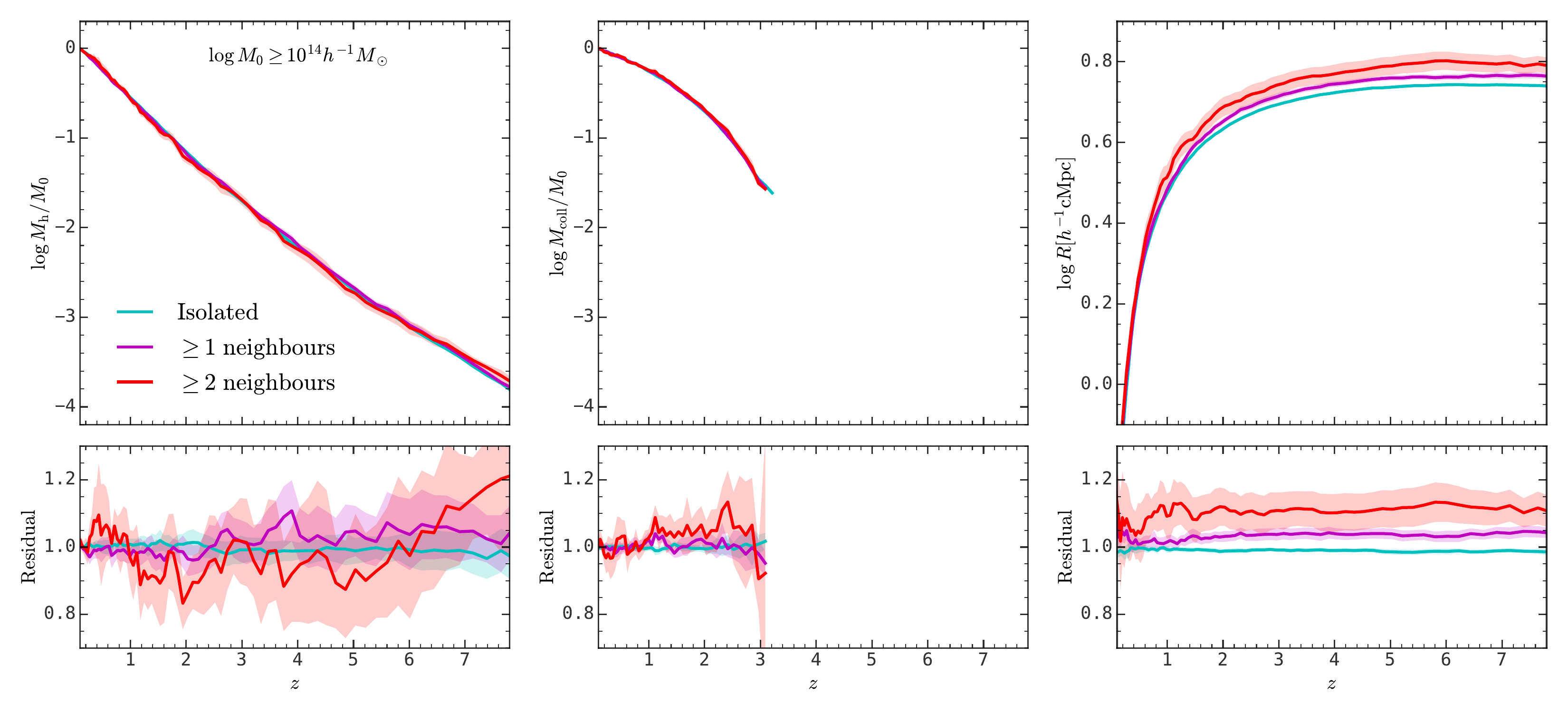}
    \caption{
        The top panels show the median mass accretion histories, collapsed mass
        history, and protohalo size histories for halos above $10^{14}h^{-1}\rm
        M_\odot$ and different numbers of neighbours within $10h^{-1}\rm Mpc$.
        The bottom panels show the fractional difference between three
        subsamples with different neighbor counts and the parent sample. The
        shaded regions show the standard deviation estimated from the bootstrap
        samples. This figure shows that the protohalo size history has a
        significant dependence on the surrounding environment of halos, but
        such dependence for the mass accretion history and the collapsed mass
        history is negligible.
    }%
    \label{fig:figures/mah_psh_on_nbr_cnt}
\end{figure*}

In addition to the auto-correlation functions shown in
Figs.~\ref{fig:figures/assembly_bias} and
\ref{fig:figures/assembly_bias_zhalf}, we also use the method in
\citet{maoAssemblyBiasExploring2018} to demonstrate the difference in the halo
assembly bias effect exhibited by the mass accretion history and the protohalo
size history for cluster-size halos above $10^{14}h^{-1}M_\odot$. To begin
with, we categorize halos into different populations according to the number of
another cluster-size halo within $10h^{-1}\rm Mpc$ of the halo in question.
Specifically, halos are defined as isolated if the neighbor counts are zero. We
also consider halos with $\geq 1$ neighbors and $\geq 2$ neighbors. We then
calculate the median mass accretion history, collapsed mass history, and
protohalo size history in each category, as presented in
Fig.~\ref{fig:figures/mah_psh_on_nbr_cnt}. Here the collapsed mass history is
defined as the total progenitor halo mass above $fM_0$ with $f=0.02$ as a
function of redshift \citep[e.g.][]{netoStatisticsCDMHalo2007,
    liHaloFormationTimes2008, gaoRedshiftDependenceStructure2008,
ludlowMassconcentrationredshiftRelationCold2016}. The fractional deviation
relative to the results obtained for the parent halo sample is presented on the
bottom panels. Note that the collapsed mass history drops below $0.02M_0$ above
$z\approx 3$. This figure demonstrates that isolated halos exhibit nearly
identical mass accretion history and collapsed mass history to the paired halos
with more than one and two neighbors \citep{liHaloFormationTimes2008}. In
contrast, paired halos exhibit larger protohalo sizes than isolated halos, and
the fractional difference is about 5\% for halos with $\geq 1$ neighbors, and
10\% for halos with $\geq 2$ neighbors.

Fig.~\ref{fig:figures/mah_psh_on_nbr_cnt} contain different information about
the halo assembly bias from Figs.~\ref{fig:figures/assembly_bias} and
\ref{fig:figures/assembly_bias_zhalf}. The auto-correlation function
characterizes the halo assembly bias on different scales, but it requires a
further compression from a linear structure of history to a scalar, which are
$z_{\rm half}$ and $\log R_c$ in this case. In contrast,
Fig.~\ref{fig:figures/mah_psh_on_nbr_cnt} compares the whole mass accretion
history and protohalo size history for paired and unpaired halos, but this
comparison is made on the chosen scale, which is $10h^{-1}\rm Mpc$ in our case.
Therefore, Fig.~\ref{fig:figures/mah_psh_on_nbr_cnt} shows that not only
$z_{\rm half}$, but also the whole mass accretion history contains no
information about the halo assembly bias.

\bsp	
\label{lastpage}
\end{document}